\documentclass[aps,prd,twocolumn,tightenlines,showpacs,amsmath,amssymb,nofootinbib]{revtex4-1}

\usepackage{graphicx}
\usepackage{amsmath,amssymb}
\usepackage{bm}
\usepackage{epsfig}
\usepackage{color}              
\usepackage{comment}
\usepackage{hyperref}
\usepackage{array, float,tabularx}

\def\be{\begin{equation}}
\def\ee{\end{equation}}
\def\ba{\begin{eqnarray}}
\def\ea{\end{eqnarray}}
\def\bal{\begin{align}}
\def\eal{\end{align}}
\def\bald{\begin{aligned}}
\def\eald{\end{aligned}}
\def\nn{\nonumber}
\newcommand{\per}{\, .}
\newcommand{\com}{\, ,}
\newcommand{\eref}[1]{Eq.~(\ref{#1})}
\newcommand{\erefs}[1]{Eqs.~(\ref{#1})}
\newcommand{\muy}{\mu_{\rm \scriptscriptstyle Y}}
\newcommand{\muw}{\mu_{\rm \scriptscriptstyle W}}
\newcommand{\uyuw}{\rm{U(1)}_{\scriptscriptstyle W} \times \rm{U(1)}_{ \scriptscriptstyle Y}}
\newcommand{\uzua}{\rm{U(1)}_{\scriptscriptstyle Z} \times \rm{U(1)}_{\scriptscriptstyle A}}
\newcommand{\ua}{\rm{U(1)}_{\scriptscriptstyle A}}
\newcommand{\uz}{\rm{U(1)}_{\scriptscriptstyle Z}}

\begin{document}

\title{Confined Vortices in Topologically Massive U(1)$\times$U(1) Theory
}

\date{\today}

\author{Mohamed M. Anber}
\email[]{mohamed.anber@epfl.ch}
\author{Yannis Burnier}
\email[]{yannis.burnier@epfl.ch}
\author{Eray Sabancilar} 
\email[]{eray.sabancilar@epfl.ch}
\author{Mikhail Shaposhnikov}
\email[]{mikhail.shaposhnikov@epfl.ch}
\affiliation{ 
Institut de Th\'eorie des Ph\'enom\`enes Physiques, EPFL, 
CH-1015 Lausanne, Switzerland.
}

\begin{abstract}
We report on a new topological vortex solution in  U(1)$\times$U(1) Maxwell-Chern-Simons theory. 
The existence of the vortex is envisaged by analytical means, and a numerical solution is obtained by integrating the equations of motion. These vortices have a long-range force because one of the U(1)s remains unbroken in the infrared, which is guarded by the Coleman-Hill theorem. The sum of the winding numbers of an ensemble of vortices has to vanish; otherwise the system would have a logarithmically divergent energy. In turn, these vortices exhibit classical confinement. We investigate the rich parameter space of the solutions, and show that one recovers the Abrikosov-Nielsen-Olesen, U(1) Maxwell-Chern-Simons, U(1) pure Chern-Simons and global vortices as various limiting cases. Unlike these limiting cases, the higher winding solutions of our vortices carry non-integer charges under the broken U(1). This is the first vortex solution exhibiting such behavior.

\end{abstract}

\maketitle

\section{Introduction}

Vortices are topological defects that were first discussed in the context of type-II superconductors by Abrikosov \cite{Abrikosov:1956sx}, where the core of a vortex is in the normal fluid phase whereas outside the core is in the superfluid phase. The relativistic generalization of vortices was given by Nielsen and Olesen \cite{Nielsen:1973cs} for the Abelian Higgs model. Vortices arise in field theories with degenerate vacuum manifolds, whose first homotopy group is non-trivial, $\pi_1[\mathcal{M}] \neq I$. According to Kibble's classification \cite{Kibble:1976sj}, e.g., the degenerate vacuum of a spontaneously broken $\rm{U(1)}$ theory has $\pi_{1} [{\rm U(1)}] = \mathbb Z$ (see e.g., Refs.~\cite{Vilenkin-Shellard,Hindmarsh:1994re} for reviews). 

The Abrikosov-Nielsen-Olesen (ANO) vortex has no electric charge, but has quantized magnetic flux, $\Phi_{B} = 2 \pi n/e$, where $e$ is the gauge coupling constant and $n \in {\mathbb Z}$ is the winding number of the Higgs field corresponding to different topological sectors classified by $\pi_{1} [{\rm U(1)}]$. As both the gauge and scalar fields are short range, they do not exhibit long range interactions. 

Interesting vortex solutions  accompany the addition of a Chern-Simons term \cite{Deser:1981wh,Deser:1982vy} $\int d^3 x ~\mu~ \epsilon^{\alpha\beta\gamma}A_{\alpha}F_{\beta\gamma}$, which breaks the $P$ and $T$ invariance of the theory and gives a mass to the photon. 
It was shown in Ref.~\cite{Paul:1986ix} that if a Chern-Simons term is added to the Abelian Higgs model, the vortices carry both a quantized magnetic flux $\Phi_{B} = 2\pi n/e$ and charge $Q = \mu \Phi_{B}$, where $\mu$ is the Chern-Simons coefficient. Similar to the ANO vortex, the interaction is short range and the charge is screened as the gauge field is higgsed (see, e.g., Refs.~\cite{Dunne:1998qy,Horvathy:2008hd} for a review of various applications of Chern-Simons vortices). 

Generally, Chern-Simons terms will appear in the context of finite temperature four-dimensional gauge theories such as the standard electroweak theory \cite{Laine:1999zi}. Upon dimensionally reducing from four to three dimensions and integrating out the fermions, non-zero Matsubara modes of the gauge bosons and the zero Matsubara mode of the temporal component of the gauge fields one obtains  Chern-Simons terms \cite{Redlich:1984md}. They are also used as effective field theory models to study the quantum Hall effect  \cite{Frohlich:1990xz,Frohlich:1991wb}.  Here, we specifically consider $\uzua$ theory with a Chern-Simons mixing term, as given by the action  (\ref{the main action of the paper}). In fact, the Chern-Simons mixing term, $\mu_1 \epsilon^{\mu \nu \alpha} \mathcal{F}_{\mu\nu} \mathcal{Z}_{\alpha}$, in (\ref{the main action of the paper}) is  the $2+1$ dimensional version of the BF theory \cite{Horowitz:1989ng}. 

In this work, we report on a new class of vortex solutions in $\uzua$ Maxwell-Chern-Simons theory. One of ${\rm U(1)}$s is spontaneously broken by a complex scalar field, whereas the other remains unbroken. As a result, the new vortex  is charged under the unbroken $\ua$, in addition of being charged under the broken $\uz$, and it mediates a long-range force.  Therefore, an ensemble of vortices and antivortices will be confined to minimize the energy of the system. This is the dynamical realization of the classical confinement that was pointed out by Cornalba and Wilczek \cite{Cornalba:1997gh} and de Wild Propitius \cite{deWildPropitius:1997wu}. Since our vortices carry magnetic fluxes, they will also exhibit non-trivial statistics in the infrared. Thus, a collection of these vortices will behave like anyons with long-range fields. We also show that the parameter space of these vortices is vast and includes the limiting cases of various known vortex solutions: ANO, U(1) Maxwell-Chern-Simons, U(1) pure Chern-Simons, and global vortices \cite{Vilenkin:1982ks}. Interestingly enough, we find that unlike these limiting cases, the $\uz$ charge and the $\ua$ magnetic flux of the higher winding solutions of our vortices are not integers times the charge and flux of the lowest winding solution. This is the first vortex solution exhibiting this behavior.

It is crucial that the model we consider does not have a self Chern-Simons term, $\mu \epsilon^{\mu\nu\beta}A_\mu F_{\nu\beta}$, for the $\ua$ gauge field which would otherwise spoil its long-range behavior. Then, one wonders if quantum corrections can generate such a term that destroys the nice long-range property of the vortices. Fortunately enough, if this term is absent on the tree and one-loop level, which is the case at hand, then the Coleman-Hill theorem \cite{Coleman:1985zi} guarantees that this term will not be generated at any higher loop level (see also Ref.~\cite{Laine:1999zi}). 

The plan of this paper is as follows. In Sec.~\ref{sec:theory}, we introduce the Chern-Simons theory with the Chern-Simons mixing term for both the mixed and unmixed basis, and then discuss their basic properties. In Sec.~\ref{sec:vortex}, we give a proof of existence for topological vorticies that are charged under the long-range $\ua$, and then present our results for the numerical solutions. In Sec.~\ref{sec:physical properties}, we calculate the flux, charge and energy of the vortex solution and of a vortex-antivortex pair. We then show that the energy of the vortex-antivortex system is finite whereas the single vortex energy is logarithmically divergent, hence the vortices are classically confined. We conclude with a summary of our results and discussion in Sec.~\ref{sec:summary and discussion}.

\section{Topologically Massive U(1)$\times$U(1) Theory}
\label{sec:theory}

We consider two topologically massive Abelian gauge fields $\mathcal{Y}_\mu$ and $\mathcal{W}_\mu$ with corresponding gauge groups $\uyuw$, and a complex scalar field $\varphi$ that is coupled to a linear combination of $\mathcal{Y}_\mu$ and $\mathcal{W}_\mu$:
\ba \bald \label{action1}
S &= \int d^3x \biggl [-\frac{1}{4} \mathcal{Y}_{\mu\nu} \mathcal{Y}^{\mu \nu} -\frac{1}{4} \mathcal{W}_{\mu\nu} \mathcal{W}^{\mu \nu}+  \muy \epsilon^{\mu \nu \alpha} \mathcal{Y}_{\mu\nu} \mathcal{Y}_{\alpha} \\ 
& \hskip 0.5cm - \muw \epsilon^{\mu \nu \alpha} \mathcal{W}_{\mu\nu} \mathcal{W}_{\alpha} + |(\partial_\mu -i g_1 \mathcal{Y}_\mu - ig_2 \mathcal{W}_\mu) \varphi|^{2}~~~~~\\
& \hskip 0.5cm - \frac{\lambda}{4} \left(|\varphi|^{2} - v^2 \right)^{2} \biggr ] \com
\eald
\ea
where $\mathcal{Y}_{\mu \nu} = \partial_\mu \mathcal{Y}_\nu - \partial_\nu \mathcal{Y}_\mu$ and $\mathcal{W}_{\mu \nu} = \partial_\mu \mathcal{W}_\nu - \partial_\nu \mathcal{W}_\mu$. For generic values of $\muy$ and $\muw$, each of the fields $\mathcal{Y}_\mu$ and $\mathcal{W}_\mu$ has a single degree of freedom which is screened in the infrared, thanks to the topological masses. Adding the two degrees of freedom of the complex scalar, our system has four degrees of freedom in total. In the following it will be useful to go to the new basis ${\cal A}_\mu$ and ${\cal Z}_\mu$:
\ba 
\bald
\mathcal{Y}_\mu  &= \cos\theta \mathcal{A}_\mu + \sin\theta \mathcal{Z}_\mu \com \\
\mathcal{W}_\mu  &= -\sin\theta \mathcal{A}_\mu + \cos\theta \mathcal{Z}_\mu \com
\eald
\ea
where $\tan\theta = g_1/g_2$. Now, we fix  
\begin{eqnarray}
\muy = \muw \tan^2\theta 
\label{choice of muy}
\end{eqnarray}
to obtain the action for the corresponding $\uzua$ theory
\ba \bald \label{action}
S &= \int d^3x \biggl [-\frac{1}{4} \mathcal{F}_{\mu\nu} \mathcal{F}^{\mu \nu} -\frac{1}{4} \mathcal{Z}_{\mu\nu} \mathcal{Z}^{\mu \nu}+ \mu_1 \epsilon^{\mu \nu \alpha} \mathcal{F}_{\mu\nu} \mathcal{Z}_{\alpha}~~~~~ \\ 
& \hskip 0.5cm + \frac{\mu_2}{2} \epsilon^{\mu \nu \alpha} \mathcal{Z}_{\mu\nu} \mathcal{Z}_{\alpha} + |D_\mu \varphi|^{2} - \frac{\lambda}{4} \left(|\varphi|^{2} - v^2 \right)^{2} \biggr ] \,,
\label{the main action of the paper}
\eald
\ea
where $\mathcal{F}_{\mu \nu} = \partial_\mu \mathcal{A}_\nu - \partial_\nu \mathcal{A}_\mu$, $\mathcal{Z}_{\mu \nu} = \partial_\mu \mathcal{Z}_\nu - \partial_\nu \mathcal{Z}_\mu$, and $D_\mu = \partial_\mu - i e \mathcal{Z}_\mu$. The parameters of the $\uzua$  theory are related to the ones in the $\uyuw$ theory as follows: $e = \sqrt{g_1^2 + g_2^2}$, $\mu_1 =  2 \muw \tan \theta$, and $\mu_2 = 2 \muw (\tan^{2}\theta-1)$. The Chern-Simons  coefficients $\mu_1$ and $\mu_2$ as well as the parameter $\lambda$ have  mass dimension $M$, while  the coupling constant $e$ and the vacuum expectation value $v$ have  mass dimension $M^{1/2}$. We set $c=1$, $\hbar =1$, $\epsilon^{012} =1$, and use the metric $\eta_{\mu\nu} = {\rm diag}(1, -1,-1)$ in what follows. 

It is a simple exercise to study the fluctuations about the vacuum $|\varphi|=v$ in Eq. (\ref{the main action of the paper}). First, the gauge field ${\cal A}_\mu$ carries a single massless degree of freedom, thanks to the unbroken $\ua$. Writing the complex field $\varphi$ as $\varphi=(v+h) e^{i\Pi}$, we find that there is a single massive radial field $h$ in the infrared.  In addition, the would-be Goldstone boson, $\Pi$, is eaten by the massive ${\cal Z}_\mu$ field. In fact, the mass of ${\cal Z}_\mu$ receives contributions from three sources: the self Chern-Simons term $\mu_2$, the Chern-Simons mixing term $\mu_1$, and the Higgs vacuum expectation value. This will be clear from our vortex solution, as is evident from Eq. (\ref{Z-boson mass}) below. Thus, the field  ${\cal Z}_\mu$ has two degrees of freedom, and we recover the total sum of the four degrees of freedom we started with. 

One wonders whether the condition (\ref{choice of muy}) and hence the spectrum described above, especially the massless $\ua$ field, are not spoiled by quantum effects. In fact, a one-loop calculation in the theory described by (\ref{action1}) does not yield any corrections to the Chern-Simons terms \cite{Laine:1999zi}. Besides, according to the Coleman-Hill theorem \cite{Coleman:1985zi}, there are no more corrections to these topological terms other than the one-loop contribution. Therefore, the massless $\ua$ gauge field is protected against quantum effects.
 
\section{Charged Vortex Solution}
\label{sec:vortex}

Throughout this work, we seek cylindrically symmetric vortex solutions of the theory given by the action (\ref{action}). 
By varying the action [\eref{action}], we obtain the field equations
\ba \bald \label{field equations}
&\partial_\beta \mathcal{F}^{\beta \sigma} + \mu_1 \epsilon^{\beta \alpha \sigma} \mathcal{Z}_{\beta \alpha} = 0 \com \\
&\partial_\beta \mathcal{Z}^{\beta \sigma} + \mu_1 \epsilon^{\beta \alpha \sigma} \mathcal{F}_{\beta \alpha} + \mu_2 \epsilon^{\beta \alpha \sigma} \mathcal{Z}_{\beta \alpha} +j^\sigma=0 \com \\ 
&D_\beta D^{\beta} \varphi + \frac{\lambda}{2} \left(|\varphi|^{2} - v^2 \right) \varphi=0 \com
\eald 
\ea
where we defined the current as
\be\label{current}
j^{\sigma} = i e \bigl [ \varphi^{*} D^{\sigma} \varphi - (D^{\sigma} \varphi)^{*} \varphi \bigr ] \per
\ee
To this end,  we take the cylindrically symmetric Nielsen-Olesen like Ans\"atze, namely,
\ba\label{ansatz}
\varphi &=& v f(r) e^{i n \theta}\,, \nn \quad 
\mathcal{Z}_i = -\epsilon^{i j} x_j \frac{Z(r)}{e r^2} \,, \\
\mathcal{Z}_0 &=& e  Z_0 (r) \,,\quad ~~~
\mathcal{A}_i = -\epsilon^{i j} x_j \frac{A(r)}{e r^2}  \,, \\
 \mathcal{A}_0 &=& e  A_0 (r) \,, \nn
\ea
where the profile functions $f(r), Z(r),  Z_0(r), A(r)$, and $ A_0(r)$ are dimensionless. Using these Ans\"atze in the equations of motion (\ref{field equations}), we obtain
\begin{eqnarray}\label{the equations of profile functs}
&&f'' + \frac{f'}{r} - (n-Z)^{2} \frac{f}{r^2} + e^4  Z_0^{2} f - \frac{\lambda v^2}{2} (f^2 -1) f = 0 \com~~ \nn\\
&&Z'' -\frac{Z'}{r} + 2e^2 v^2 f^2 (n-Z) - 2e^2 r (\mu_1  A_0' + \mu_2  Z_0') = 0 \com~~~ \nn \\
&& Z_0'' +\frac{ Z_0'}{r} - 2e^2 v^2 f^2  Z_0 - \frac{2}{e^2 r} (\mu_1 A' +\mu_2 Z') =0 \com\\
&&A'' - \frac{A'}{r} - 2 \mu_1 e^2 r  Z_0' = 0 \com \nn \\
&& A_0'' + \frac{ A_0'}{r} - \frac{2\mu_1}{e^2 r} Z' = 0 \per \nn
\end{eqnarray}
Note that in the limit $\mu_1=0$ the two U(1) sectors decouple and we obtain the equation of motion for the normal $\uz$ Chern-Simons vortex \cite{Paul:1986ix}. The last two equations in \eref{the equations of profile functs} can be integrated to find
\begin{eqnarray}
A' = 2 \mu_1 e^2 r  Z_0  \label{Aint}+{\cal D}_1 r\,,\quad
 A_0' = \frac{2 \mu_1}{e^2 r} Z +\frac{{\cal D}_2}{r} \,,\label{integration for two eqs}
\end{eqnarray}
where ${\cal D}_1$ and ${\cal D}_2$ are integration constants. In order to determine the constants ${\cal D}_1$ and ${\cal D}_2$, we examine the near core and large $r$ behavior of the system. As we shall show in Sec.~\ref{sec:boundary conditions}, the behavior of $Z(r)$ near the core goes like $r^2$, and hence, one has to set ${\cal D}_2=0$ in order to have a regular solution of the electric field $eA_{0}'(r)$ at $r=0$. Besides, a regular solution for $A'(r)$ at large $r$ demands that ${\cal D}_1=- 2 \mu_1 e^2 Z_0(\infty)$. However, since nonzero $Z_0(\infty)$ leads to a quadratically divergent energy [see \eref{hamiltonian}], it has to be set to zero, so does ${\cal D}_1$.

Before delving into the detailed vortex solution, one can read the physics of the vortex solution from the second equation in (\ref{integration for two eqs}). This relation states that starting with a single $\uz$ Chern-Simons vortex, i.e. setting $\mu_1=0$, which has an asymptotic $Z$ solution of the form $Z(\infty)= n\,, n\in {\mathbb Z}$, and  turning on a small $\mu_1$ will cause the vortex to acquire a long-range electric field proportional to $\mu_1$:
\begin{eqnarray}
E_{{\cal A}}=eA_0'\cong \frac{2\mu_1 n}{er}\,.
\end{eqnarray}
Therefore, our vortices will carry a long-range field, thanks to the unbroken $\ua$. This physics will be confirmed by detailed analytical as well as numerical checks, as we show below.

\subsection{Boundary Conditions}
\label{sec:boundary conditions}

In the usual Nielsen-Olesen vortex solution, the asymptotic behaviors of the fields are determined easily by their regularity at the core and the finiteness of the energy; namely, the profile functions vanish at the core, Higgs goes to its expectation value, whereas the gauge field goes to a pure gauge value determined by the vanishing of the kinetic energy of the Higgs field at infinity. In our model, the boundary conditions of the profile functions at $r \to \infty$ are obtained by substituting \eref{integration for two eqs} into \eref{the equations of profile functs}, setting $f=1$, and neglecting the derivative and $ {\cal O}(1/r^2)$ terms: 
\ba \bald \label{asymptotic values}
\quad Z(\infty) &= \frac{ e^2 v^2 n}{ e^2 v^2+2\mu_1^2} \,,\quad  Z_0(\infty) = 0 \, ,\quad
f(\infty) = 1\com~~~~~ \\
 A(\infty) &= n {\cal C}_2 \,,\quad  A_0(\infty) =  \frac{2 \mu_1 v^2 n}{ e^2 v^2+2\mu_1^2} \ln \frac{e^2r}{{\cal C}_1} \com
\eald
\ea
where the constants  ${\cal C}_{1,2}$ are determined numerically. Unlike the single $\uz$ Chern-Simons vortex, the ${\mathcal Z}$ winding is not an integer. This peculiar behavior will have its dramatic consequences on the single as well as the multi-vortex solutions as we discuss later on. In the limit $\mu_1 \to 0$, we find $Z(\infty) = n$ as expected for the single $\uz$ Chern-Simons vortex.

Near the core, $r\rightarrow 0$, the fields can be expanded in a Taylor series with arbitrary parameters. Requiring that the physical fields are continuous at the origin and forcing the expansion to fulfill the second order field equations (\ref{the equations of profile functs}), we can fix all but the five parameters $ a_{00},z_{00}, z_2, a_2,f_1$:
\ba \label{small r limit}
f(r)&=&f_1 r^{|n|}+\mathcal{O}(r^3)\,,\notag \\
Z(r)&=&ez_2 r^2+\mathcal{O}(r^4)\,,\notag\\
 Z_0(r)&=&\frac{z_{00}+r^2(a_2\mu_1+z_2\mu_2)}{e}+\mathcal{O}(r^4) \,,
\\
A(r)&=&ea_2 r^2+\mathcal{O}(r^4)\,,\notag\\
 A_0(r)&=&\frac{a_{00}+z_2\mu_1r^2}{e}+\mathcal{O}(r^4) \nn\,.
\ea
Note that the value of $a_{00}$ is purely a gauge choice, which doesn't contribute to the equations of motion (\ref{the equations of profile functs}). Below we will arbitrarily set $a_{00}=0$.

\subsection{Existence of the Solution}
Before moving to the numerical solution of the field equations, in this section we sketch a proof of existence of the vortex solution. To this end, we analyze the system of equations  (\ref{the equations of profile functs}) at asymptotic infinity, $r \rightarrow \infty$, taking into account the first order equations (\ref{integration for two eqs}). Since our vortices carry a long-range U(1) field, it is expected that the far-field  will follow a power-law behavior.
The asymptotic power law behavior can be obtained by expanding the profile functions as $\Psi(r) = \Psi(\infty)+\sum_{m=1}^{\infty} \psi_m /r^m$ and solving for $\psi_m$ using \eref{the equations of profile functs}. Upon performing this expansion, all constants $\psi_m$ can be fixed and we get for the asymptotic profile functions $\Psi(r) \to \Psi^{\infty}(r)$ to leading order:
\ba\label{large r large mu1}
\bald
 f^\infty(r)&= 1- \frac{4 n^2 \mu_1^4}{\lambda v^2 (e^2 v^2 + 2\mu_1^2)^2~ r^2} + {\cal O}(1/r^4) \com \\ 
 Z^\infty(r)&=Z(\infty)- \frac{16 e^2 n^3 \mu_1^6}{\lambda (e^2 v^2 + 2\mu_1^2)^4~ r^2} +{\cal O}(1/r^4)  \com \\
Z_0^\infty(r)&= - \frac{32 n^3 \mu_1^6 \mu_2 }{\lambda (e^2 v^2 + 2\mu_1^2)^5~ r^4} +{\cal O}(1/r^6) \com \\
A_0^\infty(r)&= A_0(\infty) + \frac{16 n^3 \mu_1^7}{\lambda (e^2 v^2 + 2\mu_1^2)^4 ~r^2} +{\cal O}(1/r^4) \com \\
A^\infty(r)&=n {\cal C}_2+\frac{32 e^2 n^3 \mu_1^7 \mu_2 }{\lambda (e^2 v^2 + 2\mu_1^2)^5~ r^2} +{\cal O}(1/r^4) \per
\eald
\ea
Note that in the $\mu_2=0$ limit, $Z_0(r) = A(r) = 0$ to all orders in the large $r$ expansion. In addition, in the $\mu_1 \to \infty$ limit, all the profile functions except $f(r)$ vanish identically, hence we recover the U(1) global vortex solution \cite{Vilenkin:1982ks}. 

Now we expand the profile functions around the large $r$ limit as
\ba\label{asymptotic profiles}
f &=& f^\infty+\delta f , ~~ Z=Z^\infty+\delta Z,~
Z_0=Z_0^\infty+\delta Z_0 \com~~~~
\ea
and plug these Ans\"atze back into the equations for the profile functions (\ref{the equations of profile functs}). To the leading order in the fluctuations $\delta f,\delta Z,\delta Z_0$ (and neglecting terms of the form $(\delta f,\delta Z_0,\delta Z)/r^2$), we get
\begin{eqnarray}\label{the equations with small s}
&&\delta f'' + \frac{\delta f'}{r} - \lambda v^2 \delta f \approx 0 \com~~ \nn\\
&&\delta Z'' -\frac{\delta Z'}{r} - (2e^2 v^2+4\mu_1^2) \delta Z - 2 e^2 r \mu_2 \delta Z_{0}' \approx 0 \com~~~~~~  \\
&& \delta Z_0'' +\frac{  \delta Z_0'}{r} - (2e^2 v^2+4\mu_1^2) \delta Z_0 - \frac{2}{e^2 r} \mu_2  \delta Z' \approx 0 \per \nn
\end{eqnarray}
By first writing $\delta Z(r)=e\sqrt{r}z(r)$ and $ \delta Z_0(r)=z_0(r)/(e\sqrt r)$, then solving for $z(r)$ and $z_0(r)$ at large $r$, we find following regular solutions to the homogenous equations \eref{the equations with small s}
\ba\label{large r limit}
&&\delta f(r) \simeq \frac{{\cal C}_5}{e\sqrt{r}} e^{-\sqrt{\lambda} vr} \nn\com \\
&& \delta Z(r) \simeq {\cal C}_3 e\sqrt{r} e^{-{\cal M}_Zr} \com~~~~~  \\
&&\delta Z_0(r) \simeq \frac{{\cal C}_4}{e\sqrt r} e^{-{\cal M}_Zr}\nn\,,  
\ea
where ${\cal M}_Z$ is the ${\mathcal Z}_\mu$ mass:
\begin{eqnarray}
{\cal M}_Z=-|\mu_2|+\sqrt{4\mu_1^2 +2e^2v^2+\mu_2^2} \per
 \label{Z-boson mass}
\end{eqnarray}
The solution (\ref{large r limit}) describes the intermediate region $\frac{\log(e^2v^2/2\mu_1^2)}{\sqrt{\lambda v}} \gtrsim r$, where the equality sign results by equating $f$ from the power-law behavior (\ref{large r large mu1}) to the exponential one (\ref{large r limit}).
Assuming that a vortex solution with radius $r_{\rm c}$ exists, then the profile functions and their derivatives  have to be continuous at $r_{\rm c}$. 
Thus, we match the profile functions and their derivatives in the small $r$ limit given by Eq. (\ref{small r limit}) with the large $r$ limit in Eq. (\ref{asymptotic profiles}). If the solution exists, then we should have the right number of free parameters. 
Notice that if we use the first order equations (\ref{integration for two eqs}) instead of the second order ones (\ref{the equations of profile functs}), we obtain a non-trivial relation between  $a_2$ and $z_{00}$ :
\begin{eqnarray}\label{constraint}
a_2=\mu_1 z_{00}\,,
\end{eqnarray}
which reduces the number of the free parameters by one. We are then left with three free parameters at the core $z_{00}, z_2,f_1$,   and five free constants at infinity ${\cal C}_{1,2,3,4,5}$. Now matching $f$ and its derivative across $r_{\rm c}$ gives $f_1$ and ${\cal C}_5$, $Z$ and its derivative across $r_{\rm c}$ gives $z_2$ and ${\cal C}_3$,  and $ Z_0$ and its derivative across $r_{\rm c}$ gives $z_{00}$ and ${\cal C}_4$. Matching $ A_0'$ and $A'$ across $r_{\rm c}$ does not give new information since both of these functions are dependent on $Z$ and $ Z_0$, as is clear from Eq. (\ref{integration for two eqs}). Finally, we can solve for ${\cal C}_{1,2}$ by matching $A$ and $ A_0$ across $r_{\rm c}$. 

The explicit expressions of ${\cal C}_{1,3,4,5}$ are cumbersome and not very illuminating, and we refrain from giving them here. 
It will turn out that the value of ${\cal C}_2$ determines the magnetic flux of ${\cal A}_\mu$ and partially the electric charge of ${\cal Z}_\mu$.

Now two comments are in order. First, one can read the masses of the particle spectrum from Eq. (\ref{large r limit}): the ${\cal A}_\mu$ field is massless, the Higgs mass is $\sqrt{\lambda}v$, while the ${\mathcal Z}_\mu$ mass is given by Eq. (\ref{Z-boson mass}). The $\mathcal{Z}_\mu$ mass gets contribution from the Higgs vacuum expectation value, after eating the would-be Goldstone boson, and from the topological Chern-Simons terms. Thus, as stated before, the ${\cal Z}_\mu$ field has two degrees of freedom. Second, we note that in the limit $\mu_2 = 0$, the coefficients $z_{00}$, ${\cal C}_2$ and ${\cal C}_4$ are identically zero for all values of $\mu_1$ as we checked numerically. In this limit, the Chern-Simons vortex degenerates to Abrikosov-Nielsen Olesen vortex. The vanishing of  ${\cal C}_2$ for $\mu_2=0$ means the absence of the ${\cal A}_\mu$ magnetic flux and ${\cal Z}_\mu$ electric charge as we detail below.

\subsection{Numerical Results}

\begin{figure}[t]
\begin{center}
\includegraphics[width=84mm]{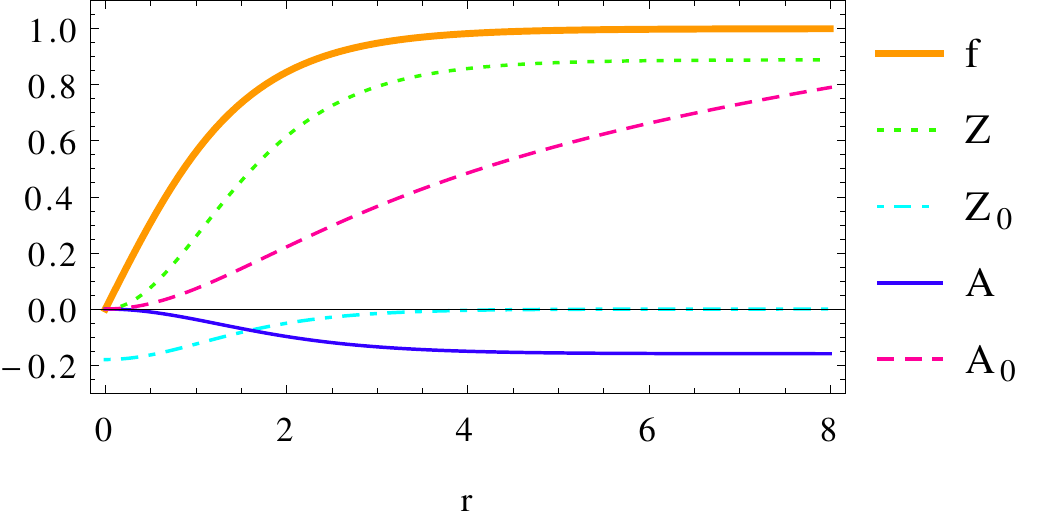}
\includegraphics[width=84mm]{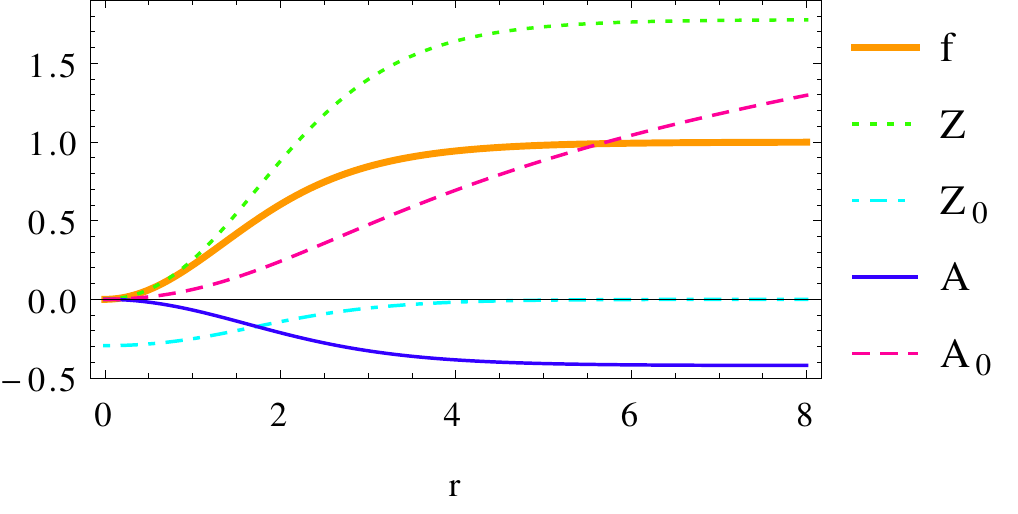}
\includegraphics[width=84mm]{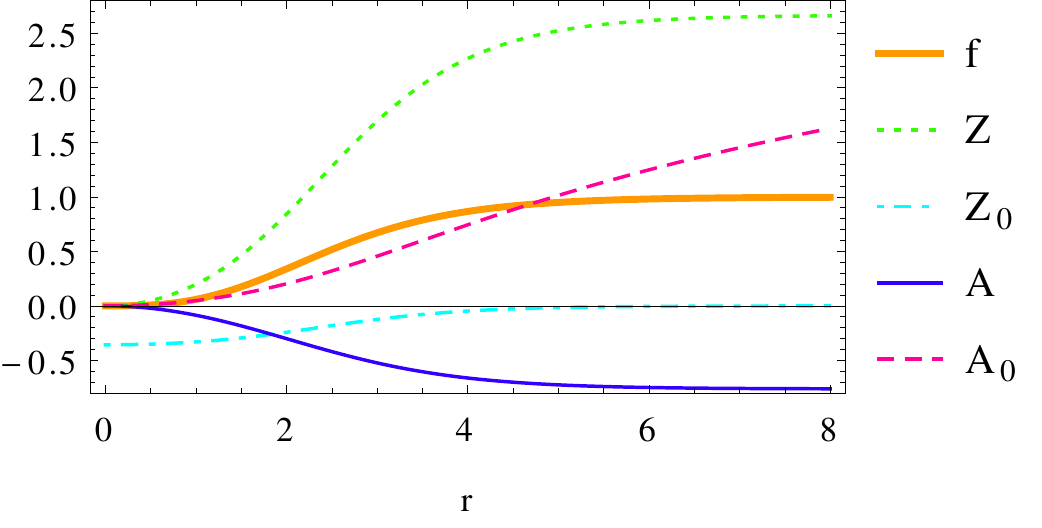}
\caption{From top to bottom, the profile functions for $n=1$, $n=2$, $n=3$ vortex solutions as a function of $r$ in units of $e^2$. Far from the core, all the fields converge as power law to a constant value except for $A_0$ which increases logarithmically. At large radius, the functions $f$, $A$, $Z$, $Z_0$, and $A_0$ tend to their asymptotic values as given by Eq. (\ref{asymptotic values}). }
\label{Fig:profile}
\end{center}
\end{figure}

\begin{figure}[t]
\begin{center}
\includegraphics[width=84.5mm]{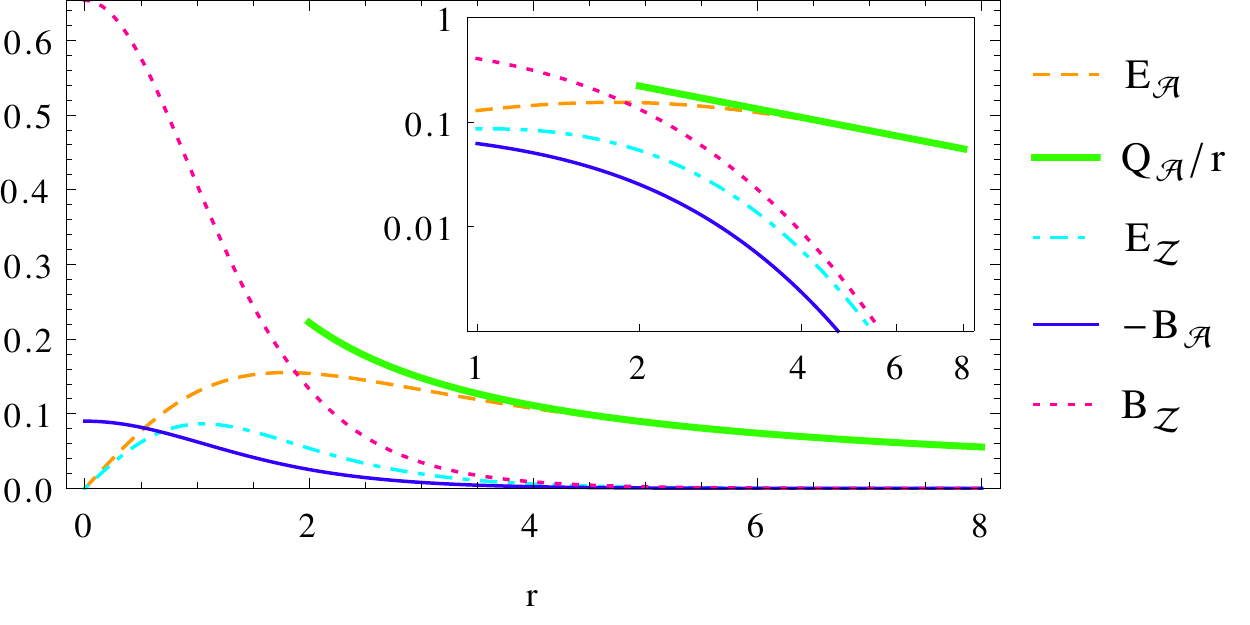}
\includegraphics[width=84.5mm]{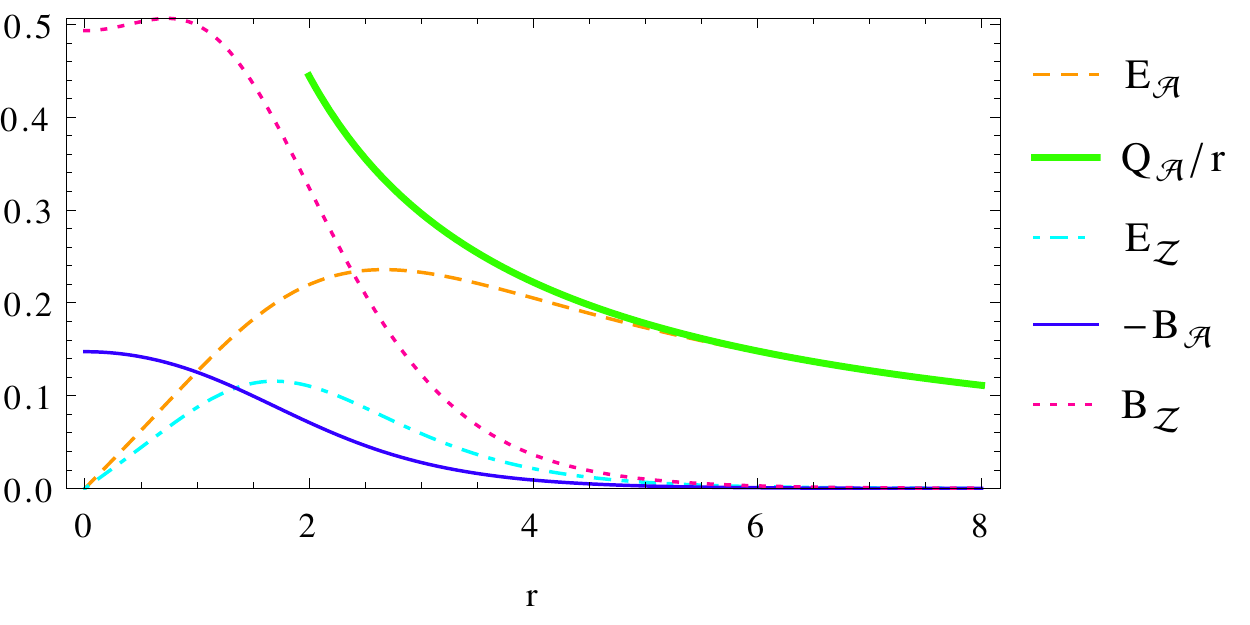}
\includegraphics[width=84.5mm]{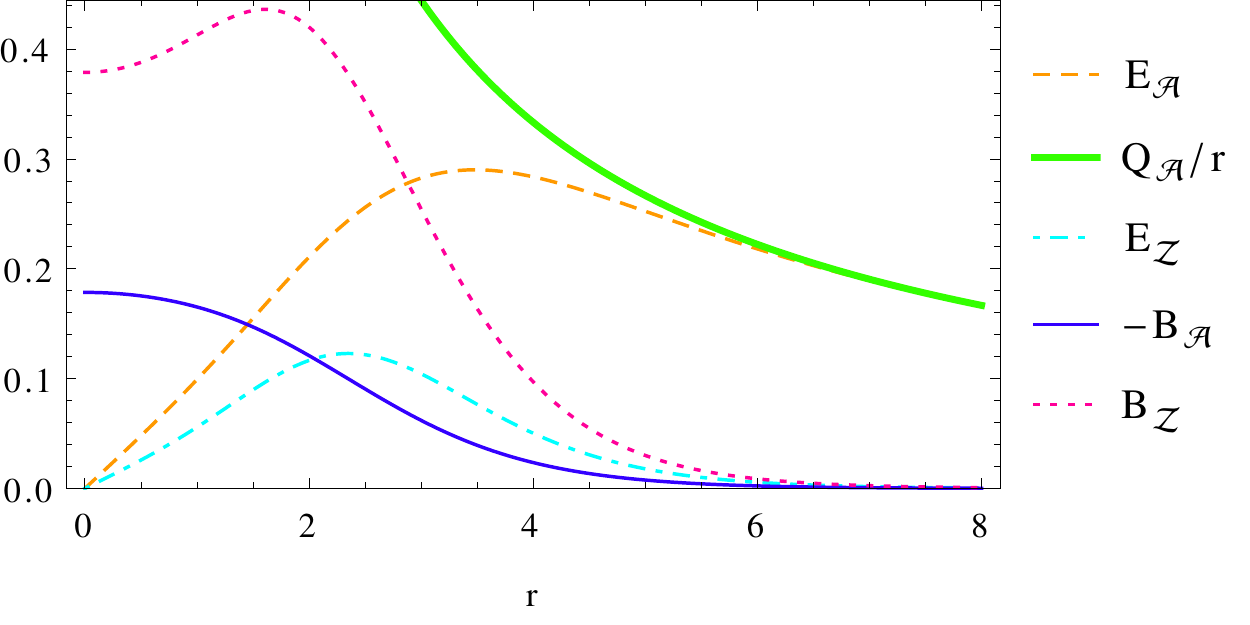}
\caption{The electric and magnetic fields as given by Eq. (\ref{elmag}) for $n=1,2,3$ from top to bottom. All fields decay as power law outside the core except for $E_{\mathcal A}$, which is a long-range field. The thick green curve is the fit to the asymptotic electric field given by $Q_{\cal A}/r$. }
\label{Fig:profile E and B fields}
\end{center}
\end{figure}

\begin{figure}[t]
\begin{center}
\includegraphics[width=84.5mm]{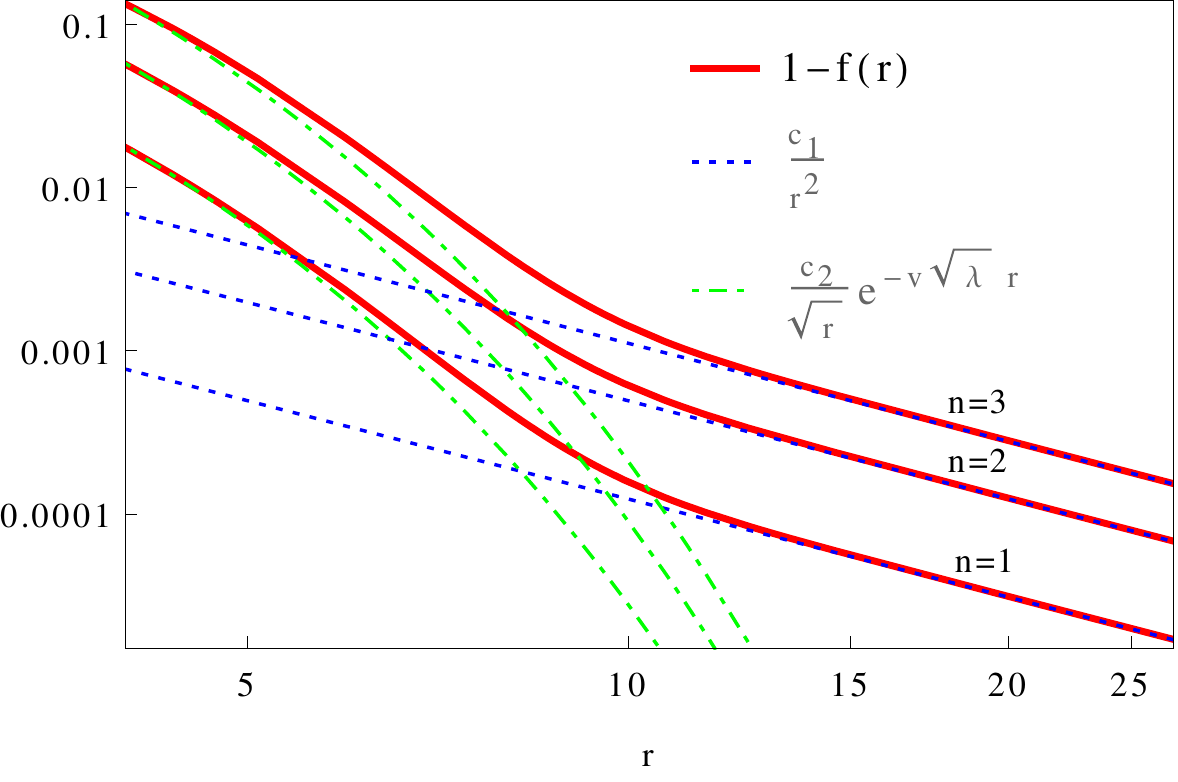}
\caption{Zooming in for the large $r$ behavior of the Higgs profile $f$ to monitor the convergence of $1-f(r)$ (red full line) towards zero. At distances larger than the vortex core, the fields first decay exponentially (the green dashed line) as in \eref{large r limit}. At larger $r$, one reaches the small power law tail (blue dotted line) of order $\mu_1^4$ as given in \eref{large r large mu1}. Note that the constant $c_2$ is fitted but the constant $c_1$ is known analytically from equation \eref{large r large mu1}.}
\label{Fig:1-f}
\end{center}
\end{figure}

\begin{figure}
\begin{center}
\includegraphics[width=80mm]{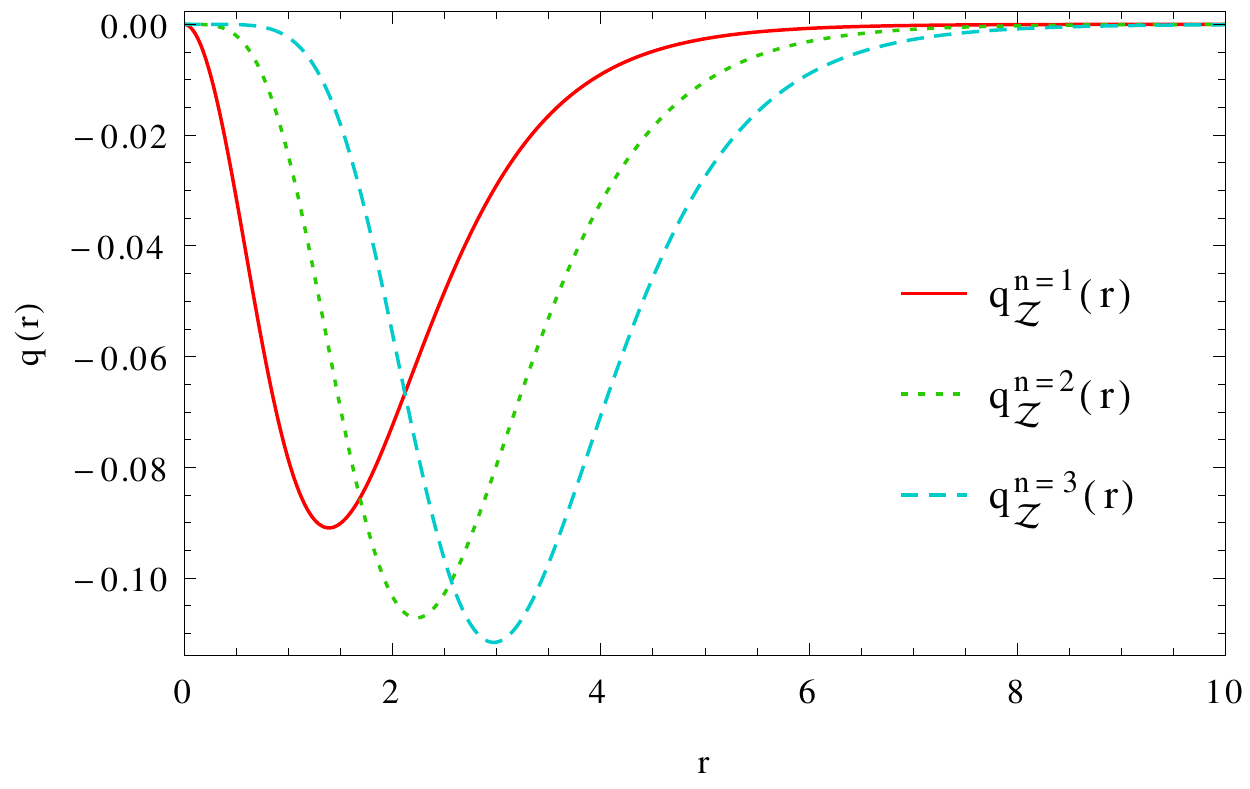}
\caption{The charge density under $\uz$ as given by the integrand in Eq. (\ref{Q}) for $n=1,2,3$. Notice that the most of the contribution to the charge comes from the near-core region. }
\label{Fig:profile charge density}
\end{center}
\end{figure}

The full numerical solution of the second order equations (\ref{the equations of profile functs}) is obtained via a shooting method as implemented in \cite{Burnier:2005he}. Starting from a small but non-vanishing radius $r_{\rm min}=10^{-5}$, in units of $e^2$, and using the small $r$ expansion (\ref{small r limit}) to sixth order, the shooting method finds the value of the four free parameters that lead to a solution satisfying the first four boundary conditions at large distance $r_{\rm max}$ as given in \eref{asymptotic values}. Here, to reach the boundary conditions to an absolute precision of $10^{-8}$ at $r_{\rm max}=27$, we use 128 digits of precision to solve the set of non-linear differential equations (\ref{the equations of profile functs}). Note that the last boundary condition at infinity in \eref{asymptotic values} is satisfied automatically. The profile functions, electric and magnetic fields, charges and energy density for the specific case $v=1$, $\lambda=1$, $\mu_1=\mu_2=1/4$ are shown in Figs.~\ref{Fig:profile}, \ref{Fig:profile E and B fields}, \ref{Fig:1-f}, \ref{Fig:profile charge density} and \ref{Fig:profile energy density} for different winding numbers, $n$, as a function of $r$, which is in units of $e^2$.

The existence of a well defined solution is ensured by checking that it follows the expansion at small $r$ given in \eref{small r limit} and merges smoothly to the asymptotic behavior given in \eref{large r large mu1}. In Fig.~\ref{Fig:1-f}, we show how the profile function $f(r)$ converges towards $1$ at large distance and see that it satisfies the large $r$ expansion [\eref{large r large mu1}]. As a further test we checked that the solution does not depend on the value of $r_{\rm min}$ and $r_{\rm max}$ as long as they remain small and large enough, respectively.
\section{Physical Properties of Charged Vortices}
\label{sec:physical properties}

\subsection{Magnetic Fluxes and Charges}
\label{sec:flux-charge}
The electric and magnetic fields as well as the kinetic term for the Higgs field can be related to the profile functions using  \eref{ansatz}:
\ba \label{elmag}
E_\mathcal{Z} &=& e  Z_0' \com \qquad B_{\mathcal{Z}} = \frac{1}{2} \epsilon^{0ij}Z_{ij} = \frac{Z'}{e r} \com \nn \\
E_\mathcal{A} &=& e  A_0' \com \qquad B_{\mathcal{A}} = \frac{1}{2} \epsilon^{0ij}F_{ij} = \frac{A'}{e r} \com \\
\mathcal{Z}_{0} &=& e Z_0 \com \qquad |{\bf D} \varphi|^2 = v^2 f'^2 + v^2 (n-Z)^2 \frac{f^2}{r^2} \,.\nn
\ea

The magnetic flux of $\mathcal{Z}_\mu$ and $\mathcal{A}_\mu$ fields for the charged vortex solution are
\ba \bald
\Phi_{B_\mathcal{Z}} &=& \oint_{S^1_{\infty}} {\bf \mathcal{Z}} \cdot {\bf d\ell} = \oint_{S^1_{\infty}} \frac{Z(r)}{er} r d \theta = \frac{2 \pi}{e} Z (\infty) \com \\
\Phi_{B_\mathcal{A}} &=& \oint_{S^1_{\infty}} {\bf \mathcal{A}} \cdot {\bf d\ell} = \oint_{S^1_{\infty}} \frac{A(r)}{er} r d \theta =\frac{2 \pi}{e} A (\infty) \,,
\eald ~~~~~~~
\ea
where $S^1_{\infty}$ is a circle enclosing the vortex at infinity.
Using \eref{asymptotic values}, we find
\ba
\Phi_{B_\mathcal{Z}} = \frac{2 \pi n}{e} \frac{ e^2 v^2 }{ e^2 v^2+2\mu_1^2} \label{phiBz} \,,\quad
\Phi_{B_\mathcal{A}} =  \frac{2 \pi n }{e} {\cal C}_2 \per
\label{the magnetic fluxes}
\ea
Although we have used the asymptotic values of the fields to calculate the fluxes, it should be clear that these fluxes originate from the near-core region of the vortex since both ${\cal Z}_\mu$ and ${\cal A}_\mu$ are screened outside the core, as is clear from Eqs.~(\ref{large r limit}) and (\ref{large r large mu1}) and Fig.~\ref{Fig:profile}. As we discussed before, ${\cal C}_2$ and hence $\Phi_{B_\mathcal{A}}$ vanish identically at $\mu_2=0$ for all values of $\mu_1$. 
Note that in the $\mu_1 \to 0$ limit \eref{phiBz} reduces to $\Phi_{B_{\mathcal{Z}}} = 2 \pi n/e$  as expected for a single $\uz$ Chern-Simons vortex \cite{Paul:1986ix}. Unlike the single $\uz$ Chern-Simons vortex, the flux in our case is not an integer times $2\pi/e$, which is attributed to the mismatch between the ${\cal Z}_\mu$ and $\varphi$ windings. More on this point will be discussed in Sec.~\ref{sec:energy}.

The charge of the vortex  under the $\mathcal{Z}_\mu$ field can be obtained from Eq. (\ref{current}) as (this is the Noether's charge)
\be\label{Q}
Q_\mathcal{Z} = \int d^2 x j^{0}  = 2 e^3 v^2 \int_{0}^{2\pi} d\alpha \int_{0}^{\infty} dr~ r f^2  Z_0 \per  
\ee 
Using the third equation in (\ref{the equations of profile functs}), we have
\ba \bald
Q_\mathcal{Z} &= 2\pi e \int_{0}^{\infty} dr \left[ (r  Z_0')' - \frac{2}{e^2} (\mu_1 A' +\mu_2 Z') \right]~~~~ \\
&=2\pi e \left[ (r Z_{0}')  - \frac{2}{e^2} (\mu_1 A +\mu_2 Z) \right]_{0}^{\infty} \com
\eald 
\ea
and then, using the boundary conditions in \eref{asymptotic values}, $Q_\mathcal{Z}$ reduces to
\be
Q_\mathcal{Z} = - \frac{4 \pi n}{e} \left[\mu_1 {\cal C}_2 + \mu_2 \frac{ e^2 v^2 }{ e^2 v^2+2\mu_1^2} \right]\per
\label{Z electric charge}
\ee
Remembering that ${\cal C}_2=0$ at $\mu_2=0$, we see right away that the Noether charge of the vortex under the ${\cal Z}_\mu$ field vanishes in this limit. In fact, the absence of the ${\cal Z}_\mu$ electric charge and ${\cal A}_\mu$ magnetic flux at $\mu_2=0$ is not a coincidence. We can understand this observation as follows.  For $\mu_1=\mu_2=0$,  we recover the normal Abrikosov-Nielsen-Olesen vortex which carries only ${\cal Z}_\mu$ magnetic flux. Turning on a non-zero value for $\mu_2$, keeping $\mu_1=0$, the Chern-Simons term will induce an electric charge for the ${\cal Z}_\mu$ field. Recalling Eq. (\ref{Q}) ---which determines the electric charge as a function of $Z_0$--- and the discussion after \eref{large r limit}, the values of the coefficients $z_{00}$ and ${\cal C}_4$ are determined upon matching the near-core and the far-region values of the profile function $Z_0$ and its derivative across the vortex wall $r_{\rm c}$. For $\mu_2 \neq 0$ both ${\cal C}_2$ and $z_{00}$ are non-vanishing, and hence we obtain non-zero values for $Q_{\mathcal Z}$  as easily  seen by taking the $\mu_1 \to 0$ limit in Eq. (\ref{Z electric charge}), which reduces $Q_{\mathcal Z}$ to $-4\pi n \mu_2/e$ as expected for the single $\uz$ Chern-Simons vortex. 

Now, let us turn on a non-zero value for $\mu_1$ keeping 
$\mu_2 \neq0$. The first equation in (\ref{integration for two eqs}) relates $A'$ to $Z_0$ at all values of $r$. At the core, non-zero values of $Z_0$, which are expected for $\mu_2 \neq 0$, will induce non-zero value for the $A$ profile which will induce magnetic field $B_{{\cal A}}$, and hence magnetic flux $\Phi_{B_{\cal A}}$. Thus, we see that the vanishing of $\mu_2$ means the vanishing of $Z_0$, and hence $Q_{\mathcal Z}$, and in sequence the vanishing of the flux $\Phi_{B_{\cal A}}$.
The interesting feature of $Q_\mathcal{Z}$ is that it is not quantized. Although the charge neutrality condition is always satisfied, i.e., $Q_\mathcal{Z}(-n)=-Q_\mathcal{Z}(n)$, we find that $Q_\mathcal{Z}(n)\neq n Q_\mathcal{Z}(n=1)$ if $\mu_1\neq 0$ and $\mu_2\neq 0$. If $\mu_2=0$ then $Q_\mathcal{Z}$ vanishes, and if $\mu_1=0$ the charge is quantized as is obvious from formula (\ref{Z electric charge}). In fact, the second term of Eq.~(\ref{Z electric charge}) is directly proportional to $n$ but the first one is not as the asymptotic value ${\cal C}_2$ depends non-trivially on $|n|$. This is the first example of a vortex with such behavior.
The $\mathcal{Z}$ charge of the vortices with winding $n=1,2,3$ are given in Table \ref{tab:1} for $\mu_2=1/4, e=v=\lambda=1$ and several values of $\mu_1$.
\begin{table}[h]
\begin{center}
\begin{tabular}{|c|c|c|c|}
\hline
	  &n=1 & n=2 & n=3\\ \hline
	$\mu_1=0~~~~$ & 3.1415 & 6.2831 & 9.4247\\ 
	$\mu_1=0.25$ & 2.2908 & 4.2625 & 5.9838  \\ 
	$\mu_1=0.50$ & 1.0752 & 1.8116 & 2.4765\\
	$\mu_1=1~~~$ & 0.2091 & 0.4066 & 0.5672 \\ 
	\hline
\end{tabular}
\end{center}
\caption{Charge $Q_\mathcal{Z}$ for $\mu_2=1/4, e=\lambda=v=1$ and different values of $\mu_1$ and winding numbers $n$. Note that it is quantized for $\mu_1=0$ (local $\uz$ vortex) and vanishes for $\mu_1\to\infty$ (global vortex), but in general this is not the case. The error estimate of the charges is $\lesssim0.01\%$.}
\label{tab:1}
\end{table}

Far in the infrared, the $\mathcal{Z}_\mu$ field is screened while the ${\cal A}_0$ field is long-range, thanks to the unbroken $\ua$. The charge of the vortex under $\ua$ can be obtained by integrating the electric field over the $\mathbb R^2$ plane and using Stokes' theorem.
Substituting the asymptotic value of  $ \mathcal{A}_0$ in \eref{asymptotic values} we obtain
\be \label{qa}
Q_\mathcal{A} =\oint_{S^1_\infty} {\bf E_\mathcal{A}} \cdot d {\bm \ell} = e\oint_{S^1_\infty}  A_0' r d\theta= \frac{4 \pi n e v^2 \mu_1}{ e^2 v^2+2\mu_1^2} \,.
\ee
This is the electric charge of the vortex under the $\ua$ field as defined from Gauss's law [note that $Q_\mathcal{A} = 2\mu_1 \Phi_{B_{\cal Z}}$, where $\Phi_{B_{\cal Z}}$ given by \eref{the magnetic fluxes}]. Therefore, an external probe with test charge $Q_{\rm test}$ will experience a force\footnote{The interaction between vortices will be elucidated in Sec.~\ref{sec:interaction}.}  $Q_{\rm test} Q_\mathcal{A}/r$. The charge $Q_\mathcal{A}$ is zero at $\mu_1=0$, increases to a maximum value of $2 \pi n/e$ at $\mu_1^2=e^2v^2/2$, and then decreases as $2\pi n e v^2/\mu_1$ for  $2 \mu_1^2 \gg e^2v^2$. The decrease of the electric charge for large values of $\mu_1$ can be understood from the equations of motion (\ref{field equations}). For large values of $\mu_1$ we can neglect the kinetic terms compared to the topological one. Since the kinetic terms are responsible for mediating the long-range $\ua$ force, we expect this force to be suppressed for small values of the kinetic terms (see also Fig.~\ref{fig:parameter space}). 

Before ending this section, let us also note that if the $\cal Z_\mu$ charge were defined from Gauss's law, as we did for $Q_\mathcal{A}$, we would find zero $\cal Z_\mu$ charge since the $\cal Z_\mu$ field is screened. 

\subsection{Energy of the Vortex}
\label{sec:energy}

The energy of the vortex can be calculated starting from the Hamiltonian density of the theory defined by \eref{action}. Because the vortex is static, one can alternatively use the Euclidean version of this action. In fact, since the Chern-Simons terms do not depend on any background metric, these terms do not contribute to the energy-momentum tensor and hence to the Hamiltonian. In turn, one does not expect that both the Hamiltonian and the Euclidean action to have the same functional form. Irrespectively, we checked that both formulations give the exact same answer for the vortex energy. 

\begin{figure}[t]
\begin{center}
\includegraphics[width=80mm]{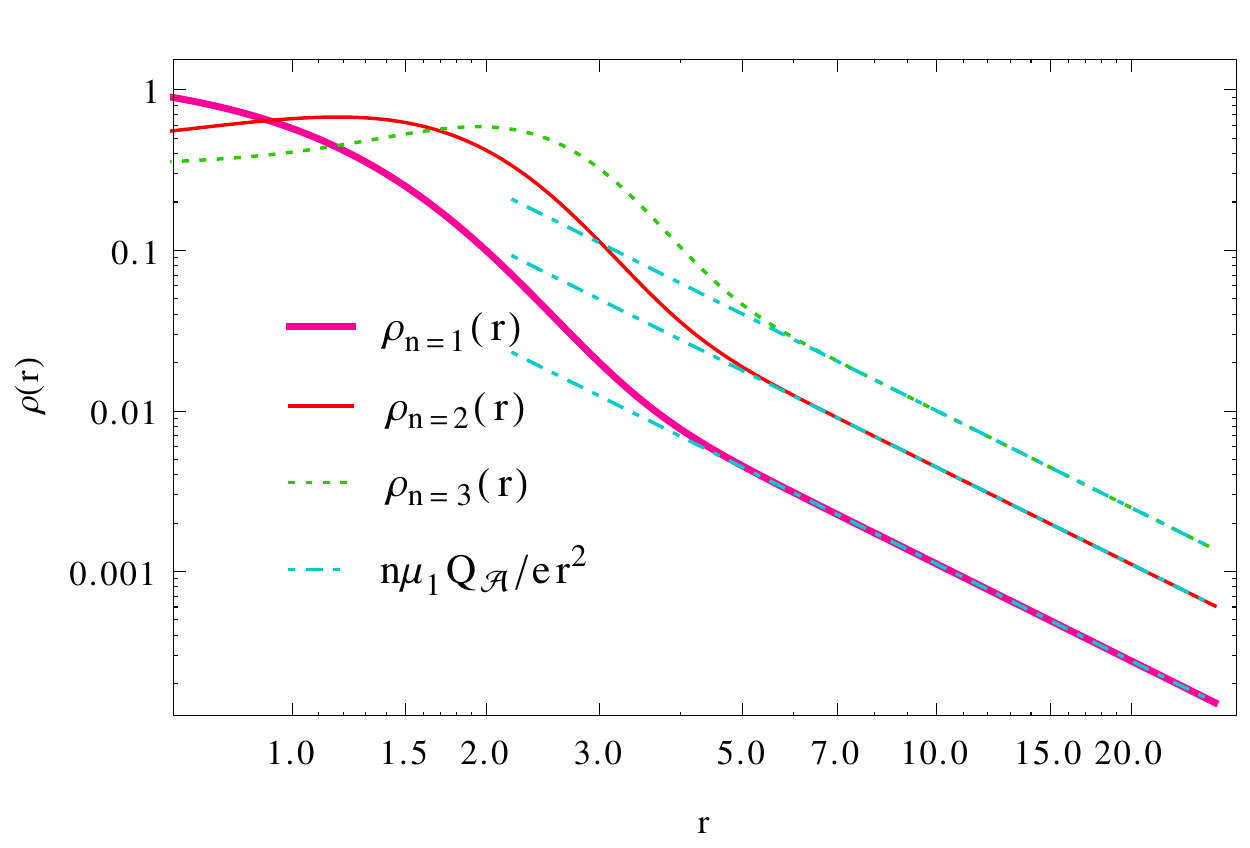}
\caption{The energy density of a vortex with $n=1,2,3$. The blue dash-dotted lines show the asymptotic behavior given by the derivative of the infrared contribution to the energy given by \eref{energy ir} with respect to $r$. }
\label{Fig:profile energy density}
\end{center}
\end{figure}

The Hamiltonian density  is given by\footnote{Note that our formula differs from the one given in Ref.~\cite{Horvathy:2008hd} by a sign of the $e^4  Z_0^2 |\varphi|^2$ term, which should be positive. Besides, the Chern-Simons term that they have in the Hamiltonian density should not be included as it does not contribute to the energy momentum tensor.}
\ba \bald \label{Hv}
\mathcal{H}_{\rm v} &= \frac{1}{2} \left( E_{\mathcal{Z}}^{2} + B_{\mathcal{Z}}^{2} + E_{\mathcal{A}}^{2} + B_{\mathcal{A}}^{2} \right) + e^4  Z_0^2 |\varphi|^2 \\
&\hskip 0.5cm + |{\bf D} \varphi|^2 + \frac{\lambda}{4} (|\varphi|^{2} - v^2)^{2}\,.
\eald
\ea
Next, we split the total Hamiltonian density into three parts: the core $\mathcal{H}_{\rm c}$, electric $\mathcal{H}_{E_{\mathcal{A}}}$ and Goldstone $\mathcal{H}_{\rm G}$. Using \eref{elmag},
\ba\label{hamiltonian}
\nonumber
\mathcal{H}_{\rm c} &=&\frac{1}{2}\biggl [ e^2  Z_0'^2 + \frac{Z'^2}{e^2 r^2} +\frac{A'^2}{e^2 r^2} + 2 e^4 v^2  Z_0^2 f^2~~~~~~~\\
&& \hskip 0.5cm + 2 v^2 f'^2 \nn+ \frac{\lambda v^4}{2} (f^{2} - 1)^{2}  \biggr] \com \\
\label{different parts of energy}
\mathcal{H}_{E_{\mathcal{A}}} &=& \frac{e^2}{2}  A_0'^2 \com \\
\nonumber
\mathcal{H}_{\rm G} &=& v^2 (n-Z)^2 \frac{f^2}{r^2} \per
\ea
Integrating each term over $\mathbb{R}^2$, the total energy of a vortex is obtained as
\ba \label{Ev}
\mathcal{E}_{\rm v} &=& \mathcal{E}_{\rm c}+\mathcal{E}_{E_{\mathcal{A}}} +\mathcal{E}_{\rm G} \per
\ea
The core contribution to the energy, $\mathcal{E}_{\rm c}$ can be found numerically (see Fig.~\ref{Fig:profile energy density}),
while in the IR only the electric $\mathcal{E}_{E_{\mathcal{A}}}$ and Goldstone $\mathcal{E}_{\rm G}$ contributions remain, and the energy can be calculated analytically:
\ba \label{energy ir}
\mathcal{E}_{\rm \scriptscriptstyle IR} &\approx& \mathcal{E}_{E_{\mathcal{A}}}^{\rm \scriptscriptstyle IR}+ \mathcal{E}_{\rm G}^{\rm \scriptscriptstyle IR}  \nn \\
&=&  \pi \biggl [ \frac{4 \mu_1^2}{e^2} Z(\infty)^2 + 2 v^2 (n-Z(\infty))^2 \biggr] \int_{r_{\rm c}}^{R_{\rm \scriptscriptstyle IR}} d \ln r  ~~~~~~~\nn \\
&=&  \frac{4 \pi n^2 v^2 \mu_1^2}{ v^2 e^2 + 2 \mu_1^2} \ln \frac{{R_{\rm \scriptscriptstyle IR}}}{r_{\rm c}} \com
\ea
where we imposed an IR cutoff at $R_{\rm \scriptscriptstyle IR}$. Exactly like an electric charge in 2+1 dimensions, the energy of a single vortex is logarithmically divergent. Note that there are two contributions to the energy. The first one comes form the electric field $\mathcal{E}_{E_{\mathcal{A}}}$, which is expected from the long-range $\ua$ outside the vortex core. The second contribution, the Goldstone energy, $\mathcal{E}_{\rm G}$ is more interesting. In a trivial background, $\varphi=0$, we can study the spectrum of the symmetry breaking of a global ${\rm U(1)}$ symmetry by writing $\varphi(x)=\left[v+h(x) \right]e^{i\Pi(x)}$, where $h$ is the Higgs boson and the phase $\Pi$ is the Goldstone boson. This Goldstone boson is a physical massless degree of freedom that exists in the spectrum of the theory. However, once we gauge ${\rm U(1)}$ this Goldstone will be eaten by the corresponding gauge field which in turn acquires a mass\footnote{This can be seen once we compute the square of the covariant derivative $|D_\mu \varphi|^2$ which gives $v^2(\partial_\mu\Pi-e{\cal Z}^f_\mu)^2$, where ${\cal Z}_\mu^f$ is the fluctuation field. Using the gauge transformation ${\cal Z}_\mu\rightarrow {\cal Z}_\mu+\partial_\mu\Pi/e$ (unitary gauge), we immediately recognize $2e^2v^2$ as the {\cal Z}-mass.}. In fact, the physical spectrum does not contain any Goldstone boson since there is no symmetry to break: the gauged ${\rm U(1)}$ is a redundancy rather than a genuine symmetry. If the background is non-trivial, as in our vortex case, then one has to repeat the same argument in the given background. In this case, we write $\varphi(x)=v \left[f(r)+h(x)\right]e^{i\left[n\theta(r)+i\Pi(x)\right]}$, where $f(r)$ is the profile function of the vortex. We also write the field ${\cal Z}_\mu$ as ${\cal Z}_\mu(x)={\cal Z}_{\mu}^b(x)+{\cal Z}_\mu^f(x)$, where the background solution ${\cal Z}_{\mu}^b$ can be read directly from the profile functions $Z(r)$ and $Z_0(r)$ given above. Now the square of the covariant derivative gives $v^2 \left[\partial_\mu(n\theta+\Pi)-e({\cal Z}_{\mu}^b+{\cal Z}_\mu^f)\right]^2$. Again, one can use an appropriate gauge transformation to kill the fluctuating Goldstone field $\Pi$, thus interpreting $2e^2v^2$ as the mass of ${\cal Z}_\mu$ in our non-trivial background. What remains is the {\em Goldstone background} contribution $v^2 \left(n\partial_\mu\theta-e{\cal Z}_{\mu}^b\right)^2$ which gives $\mathcal{H}_{\rm G}$ in Eq. (\ref{different parts of energy}). As we showed above,  this Goldstone background energy is logarithmically divergent\footnote{In fact, one can obtain the same behavior in the presence of a non-dynamical static magnetic $B_{\cal Z}$ field with non-integer flux. In this case, we also find that the {\it Goldstone background} energy is logarithmically divergent because of the mismatch between the winding of the Higgs and gauge fields. We thank T. Sulejmanpasic for emphasizing this point.}. Therefore, unlike the Abrikosov-Nielsen-Olesen vortices or their single ${\rm U(1)}$ Chern-Simons cousins, our vortices are not genuine solitons. This  forces us to consider an ensemble of an equal number of vortices and anti-vortices. These in turn will form vortex-antivortex confined pairs that lowers the total energy of the system. The next section is devoted  to the study of the interaction energy of the vortex-antivortex pair.

\subsection{Interaction between Two Vortices}
\label{sec:interaction}

In this section, we show that the total energy of a system of a vortex and an antivortex pair is finite. Treating the 2-vortex system is rather complicated. However, it will be sufficient to approximate the system as a superposition of a pair of vortices with opposite winding numbers located at a large separation $R$. As long as the separation is large enough, the individual solutions do not receive a considerable modification by the presence of the other vortex. With these assumptions, the total scalar and gauge fields of a vortex-antivortex system at $|{\bf x}|\gg r_{\rm c}$ can be approximated as follows:
\ba \bald \label{2vortex}
\varphi &\cong v e^{i n \theta_1({\bf x}- {\bf x_1}) - i n \theta_2({\bf x}-{\bf x_2})} \com  \\
\mathcal{Z}_0 &\cong 0 \com \\
\mathcal{Z}_i &\cong -\frac{Q_\mathcal{A}}{4 \pi \mu_1} \epsilon_{i j} \left[ \frac{({\bf x} - {\bf x_1})_j}{|{\bf x} - {\bf x_1}|^2} - \frac{({\bf x} - {\bf x_2})_j}{|{\bf x} - {\bf x_2}|^2}  \right] \com \\
\mathcal{A}_0 &\cong \frac{Q_\mathcal{A}}{2\pi} {\rm ln} \frac{|{\bf x} - {\bf x_1}|}{|{\bf x} - {\bf x_2}|} \com  \\
\mathcal{A}_i &\cong -\frac{n \mathcal{C}_2}{e} \epsilon_{i j} \left[ \frac{({\bf x} - {\bf x_1})_j}{|{\bf x} - {\bf x_1}|^2} - \frac{({\bf x} - {\bf x_2})_j}{|{\bf x} - {\bf x_2}|^2}  \right] \com
\eald
\ea
where, $Q_{\mathcal{A}}$ is given by \eref{qa}, ${\bf x_1}$ and ${\bf x_2}$ are the locations of the vortex and antivortex, respectively. Upon substituting \erefs{2vortex} in \eref{Hv}, the total Hamiltonian density in the IR can be obtained as:
\ba \label{Hir}
\mathcal{H}_{\rm \scriptscriptstyle IR} (R, r,\alpha) = \frac{8 n^2 v^2 \mu_1^2}{e^2 v^2 + 2 \mu_1^2} \frac{(R/2)^2}{r_1^2 r_2^2} \com
\ea
where we defined $r \equiv |{\bf x}|$,
\ba \bald
r_1 \equiv \sqrt{r^2 - R r \cos\alpha + (R/2)^2} ~\com \\
r_2 \equiv \sqrt{r^2 + R r \cos\alpha + (R/2)^2}~ \,,
\eald
\ea
$R$ is the separation distance between the two vortices, and $\alpha$ is the polar angle.
Then, the total energy can be found upon integrating $\mathcal{H}_{\rm \scriptscriptstyle IR} (r,\alpha)$ over the polar angle $\alpha$ and radial variable $r$: 
\ba \bald \label{Hir}
\mathcal{E}_{\rm \scriptscriptstyle IR} (R) &= \int_\mathcal{C} dr~r~ \int_{0}^{2\pi} d \alpha~ \mathcal{H}_{\rm \scriptscriptstyle IR} (R, r,\alpha) \\
&= \frac{8 n^2 v^2 \mu_1^2}{e^2 v^2 + 2 \mu_1^2}~ \frac{\pi}{2} \ln \frac{R^4 + r_{\rm c}^4}{R^2 r_{\rm c}^2 - r_{\rm c}^4} \per
\eald
\ea
Since $R \gg r_{\rm c}$, the total energy of a vortex-antivortex system can be simply written as:
\ba
\mathcal{E}_{\rm \scriptscriptstyle IR} (R) \approx \frac{8 \pi n^2 v^2 \mu_1^2}{e^2 v^2 + 2 \mu_1^2}~ \ln \frac{R}{r_{\rm c}} + \mathcal{O}(r_{\rm c}/R) \per
\ea
Here, for simplicity we took the region of integration for the radial variable to be $\mathcal{C} = \{ 0 \leqslant r \leqslant R/2 - r_{\rm c} ~{\rm and}~ R/2 + r_{\rm c} \leqslant r \leqslant \infty \}$ to remove the contribution of the cores of the vortices. It would be more accurate to cut out just two discs of radii $r_{\rm c}$ centered at the cores of the vortex and antivortex. However, removing the contribution of this tiny strip of radius $2r_{\rm c}$ only leads to  an error of order $r_{\rm c}/R$, which is a subleading effect that we ignore here.

The total energy takes an even simpler form in the $2\mu_1^2 \ll e^2 v^2$ limit, i.e., when the $\mu_1$ contribution to ${\cal Z}$-mass is small compared to the mass coming from the spontaneous breaking of the $\uz$ symmetry:
\ba
\mathcal{E}_{\rm \scriptscriptstyle IR} (R)|_{2\mu_1^2 \ll e^2 v^2} \sim \frac{Q_{\mathcal{A}}^2}{2 \pi}~ \ln \frac{R}{r_{\rm c}} \per
\ea
In other words, the interaction of a vortex and an antivortex in this particular limit is exactly like that of two point particles with opposite charges in $2+1$ D. 

In the opposite limit $2\mu_1^2 \gg e^2 v^2$, the total energy takes the form
\ba
\mathcal{E}_{\rm \scriptscriptstyle IR} (R)|_{2\mu_1^2 \gg e^2 v^2} \sim 4 \pi n^2 v^2 ~ \ln \frac{R}{r_{\rm c}} \per
\ea
In this limit, $Q_{\mathcal{A}} \sim 0$, hence, the contribution is mostly due to the {\it Goldstone background}. Note that this is nothing but the energy of a global vortex-antivortex pair (see Fig.\ref{fig:parameter space}). 

To summarize, a system of a vortex-antivortex pair has a finite energy and is logarithmically confined. The total energy gets contributions both from the electric field and {\it Goldstone background} of the vortices.

\section{Summary and Discussion}
\label{sec:summary and discussion}

In this work, we have obtained a new vortex solution in the $\uzua$ Chern-Simons gauge theory. These vortices are classified by a topological number $n \in \mathbb Z$. Inside the core of the vortex the Higgs field is in its symmetric phase and both ${\rm U(1)}$s are topologically massive. Outside the core,  the Higgs field gets a vacuum expectation value causing the spontaneous breaking of $\uz$. The resulting Goldstone boson is eaten by the corresponding gauge field, namely the  ${\cal Z}_\mu$ field, which now acquires an extra degree of freedom. Thus, the mass of the ${\cal Z}$-boson gets contributions from both the topological terms and the Higgs vacuum expectation value, as is evident from Eq. (\ref{Z-boson mass}). In addition to the massive ${\cal Z}$-boson, the vortex mediates a long-range force outside its core, thanks to the unbroken $\ua$. This adds up correctly to the number of degrees of freedom (d.o.f.): inside the core the Higgs has 2 d.o.f. and each ${\rm U(1)}$ has a single d.o.f., and outside the core the massive ${\cal Z}$-boson has 2 d.o.f. while the ${\cal A}$-boson (photon) and Higgs each has a single d.o.f. 

\begin{figure}[t]
\begin{center}
\includegraphics[width=40mm]{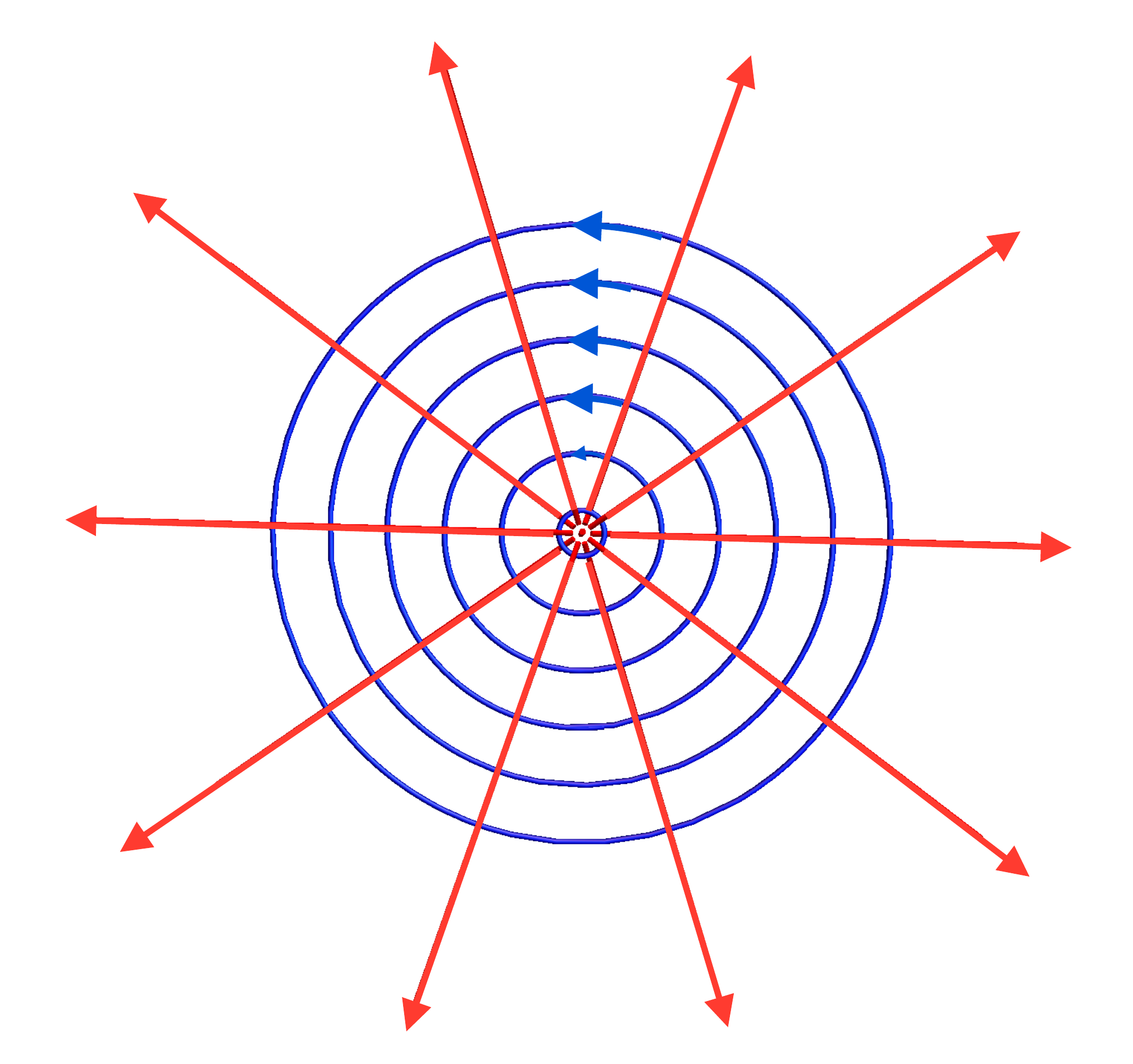}\includegraphics[width=40mm]{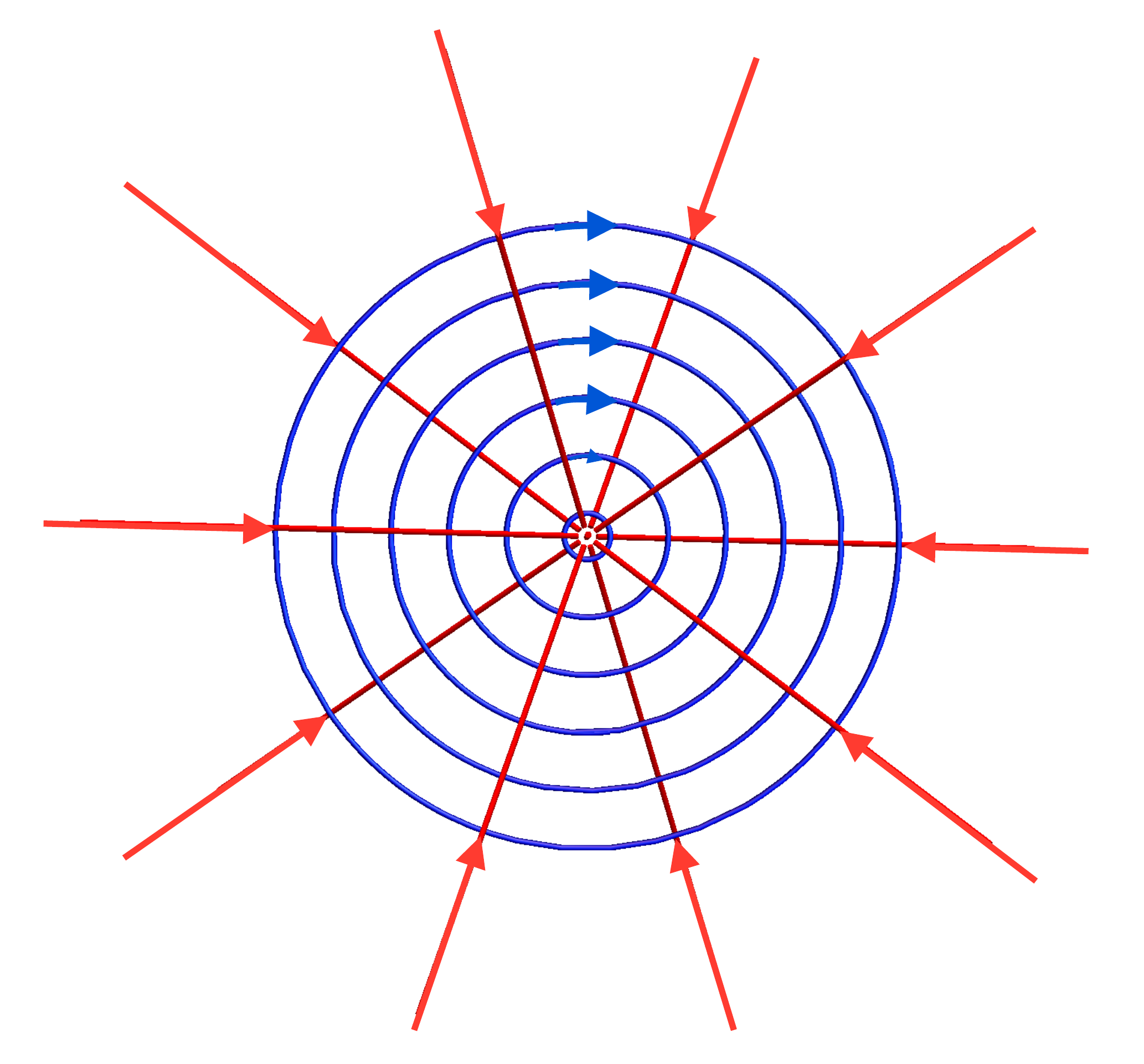}
\caption{The field lines of a vortex (left) and an antivortex (right). The radial (red) and circular (blue) field lines correspond to the electric field [\eref{EA}] and Goldstone background [\eref{goldstone field}], respectively.}
\label{fig:field lines}
\end{center}
\end{figure}

Our vortices are charged under the $\ua$ and hence two vortices will have a logarithmic interaction due to the massless $\ua$ field.  The electric field  $\bold E_{\cal A}$ is given by
\begin{eqnarray} \label{EA}
\bold E_{\cal A}=\frac{Q_{\cal A}}{r} \hat{\bm e}_r\,,\quad Q_{\cal A}= \frac{4 \pi n e v^2 \mu_1}{ e^2 v^2+2\mu_1^2} \,,
\end{eqnarray}
where $\hat{\bm e}_r$ is a unit vector in the radial direction. In addition to the long-range electric field, our vortices also interact due to a background Goldstone field. As we stressed at the end of Sec.~\ref{sec:energy}, this Goldstone background is different from the Goldstone fluctuations which are completely absorbed by the ${\cal Z}$-boson. Let us define the Goldstone field:
\begin{eqnarray}
G_i=v\partial_i \theta -{\cal Z}_i \com
\end{eqnarray}
which can be rewritten as
\begin{eqnarray}\label{goldstone field}
\bold G=\frac{Q_G}{r} \hat{\bm e}_\alpha\,,\quad  Q_G=\frac{4\sqrt{2} \pi n v \mu_1^2}{e^2v^2+2\mu_1^2}\,,
\end{eqnarray}
where $\hat{\bm e}_\alpha$ is a unit vector in the polar direction, and we have defined an effective Goldstone charge $Q_G$ (see Fig.~\ref{fig:field lines}).
 Then, the total interaction energy between vortices with charges $(Q_{\cal A}^1,G_G^1)$ and $(Q_{\cal A}^2,G_G^2)$ is given by 
\begin{eqnarray}\label{interaction equation}
\frac{1}{2\pi} \left(Q_{\mathcal{A}}^1Q_{\mathcal{A}}^{2}+Q_G^1Q_G^2\right) \ln \frac{R}{r_c}\,.
\end{eqnarray}

\begin{figure}[t]
\begin{center}
\includegraphics[width=85mm]{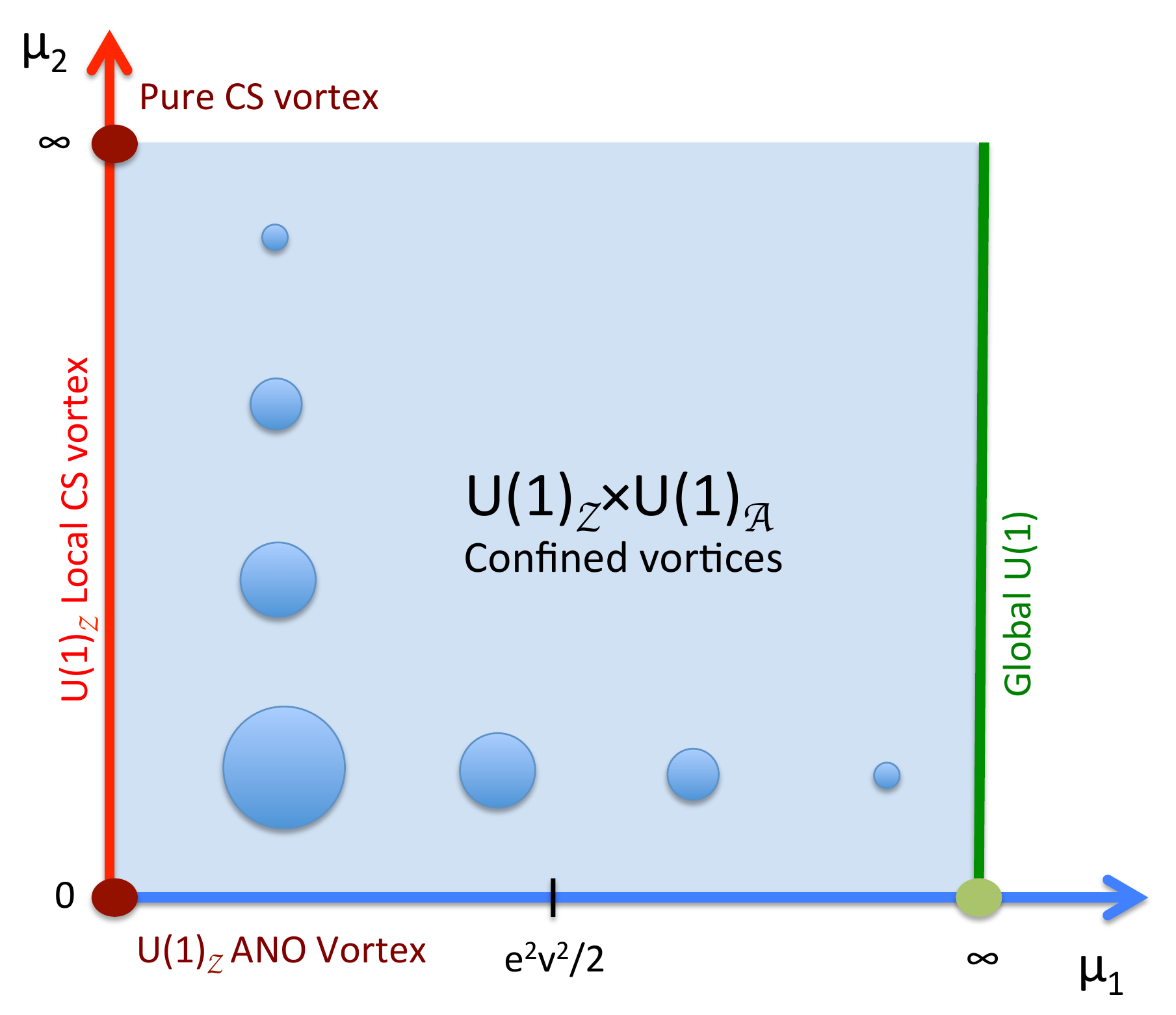}
\caption{The parameter space of the vortex solution. The circle size indicates the core radius, which decreases as we increase both $\mu_1$ and $\mu_2$. Our solution covers the whole region including the boundaries, and reduces to the known solutions in certain limits of Chern-Simons coefficients $\mu_1$ and $\mu_2$. $e^2v^2/2$ marks the boundary between electric field dominated vs. Goldstone background dominated solutions.}
\vskip 0.4cm
\label{fig:parameter space}
\end{center}
\end{figure}

\begin{figure}[t]
\begin{center}
\vskip 0.4cm
\includegraphics[width=82mm]{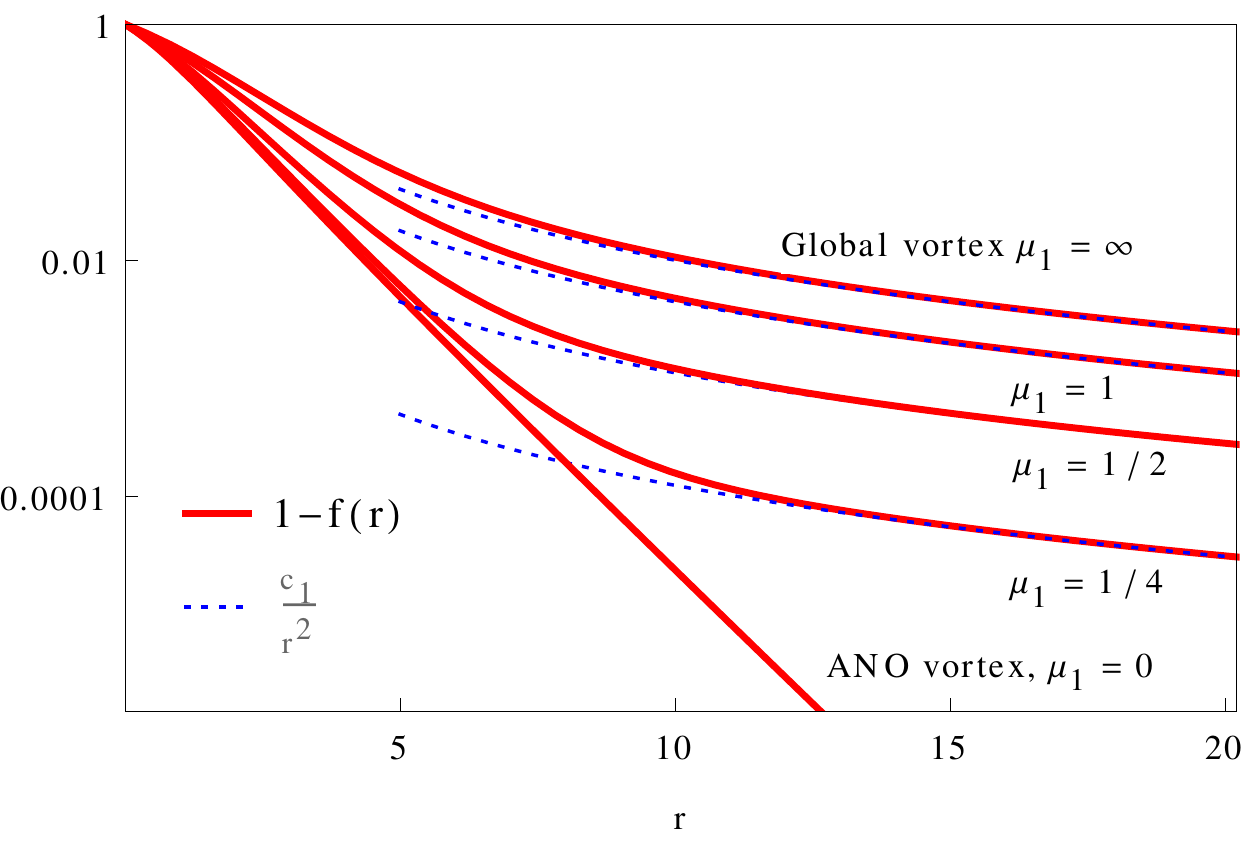}
\caption{Profile for the Higgs field $1-f(r)$ for different $\mu_1$'s ($\mu_2=0$ here). We see that the vortex profile goes exponentially to the Higgs vev (straight line on the log-linear plot) in the case of the ANO vortex ($\mu_1=0$) and continuously changes with increasing $\mu_1$ to a power law, to our vortices and to the global vortex in the limit $\mu_1=\infty$.}
\label{fig:fmu}
\end{center}
\end{figure}

At $\mu_1=\mu_2=0$  our vortices reduce to  the Abrikosov-Nielsen-Olesen vortices. Turning on a non zero value of $\mu_2$, but still setting $\mu_1=0$,   we recover the single $\uz$ Chern-Simons vortex. For both cases the charges $Q_G$ and $Q_{\cal A}$ are zero, and the vortices do not interact with long-range forces. As we turn on a non-zero value for $\mu_1$, our vortices start interacting logarithmically. For values of $\mu_1^2 \ll e^2 v^2/2$ the interaction is dominated by the electric force, while the Goldstone force is subleading. This picture is reversed for $\mu_1^2 \gg e^2 v^2/2$ as the Goldstone force dominates over the electric one. Both pictures are independent of the value of $\mu_2$. The only effect of $\mu_2$ is that it contributes to the ${\cal Z}$ mass, as is clear from \eref{Z-boson mass}, making it infinitely large as  $\mu_2\rightarrow \infty$. On the other hand,  taking $\mu_1 \rightarrow \infty$, the electric charge as well as the magnetic flux $\Phi_{B_{\cal A}}$ vanish, the later is clear from the fact that all the gauge fields decouple in this limit, and the interaction is solely due to the Goldstone background. In fact, this is the limit where we recover global vortices. Of course, in this limit the ${\cal Z}$-boson becomes infinitely massive and decouples. The infinite ${\cal Z}$ mass can be thought of as a UV cutoff on the vortices. The parameter space diagram of these different limits is illustrated in Fig.~\ref{fig:parameter space}. In Fig.~\ref{fig:fmu}, we show the profile function $f$, and its asymptotic behavior for these regimes.
Since the energy of a single vortex is logarithmically divergent, its long energy tail has to be trimmed by either putting the vortex in a container with a finite radius, or by considering an equal number of vorticies and antivortices. In the later case, the system will lower its total energy by forming confined vortex-antivortex pairs.  

So far, we have not discussed the effect of the magnetic fluxes $\Phi_{B_{\cal A}}$ and $\Phi_{B_{\cal Z}}$  on the behavior of the vortices. It was shown in \cite{Bais:1993nn} that the monodromy  of a particle $(\Phi_{B_{\cal A}},\Phi_{B_{\cal Z}})$ and a remote particle $(\Phi_{B_{\cal A}}',\Phi_{B_{\cal Z}}')$ leads to an Aharonov-Bohm phase 
\begin{eqnarray}\label{AB phase}
\exp\left[i\mu_1\left( \Phi_{B_{\cal Z}}\Phi_{B_{\cal A}}'+\Phi_{B_{\cal Z}}'\Phi_{B_{\cal A}}\right)+i\mu_2\Phi_{B_{\cal Z}}\Phi_{B_{\cal Z}}' \right] \per~~
\end{eqnarray} 
This phase gives rise to  non-trivial statistics of the vortices, which behave as anyons. One can understand the origin of this phase as follows. Given a charge $q$ that couples to a vector potential ${\cal A}_\mu$, this particle acquires the Aharonov-Bohm phase  $\exp\left[ iq\oint dx^\mu {\cal A}_\mu \right]=\exp[iq\Phi_{B_{\cal A}}]$, where $\Phi_{B_{\cal A}}$ is the flux of ${\cal A}_\mu$, as the particle makes a non-contractible winding around the source ${\cal A}_\mu$. Now, the zeroth component of the equations of motion (\ref{field equations}) (the Gauss's laws) can be written in the form  
\begin{eqnarray}
\int_{\mathbb R^2} d^2x \nabla\cdot {\bold E_{\cal A}}&=&2\mu_1 \Phi_{B_{\cal Z}} \,,\label{Gausss law A}
\\
\int_{\mathbb R^2}d^2x \left[ \nabla\cdot {\bold E_{\cal Z}}+2e^2v^2 f^2 {\cal Z}_0 \right]&=&2\mu_1 \Phi_{B_{\cal A}}+2\mu_2\Phi_{B_{\cal Z}}\,.~~~~~~
\label{Gausss law Z}
\end{eqnarray}
Thus,  $\mu_1\Phi_{B_{\cal Z}}$ is an effective charge which couples to the flux of  $\cal A_\mu$, while $\mu_1 \Phi_{B_{\cal A}}+\mu_2\Phi_{B_{\cal Z}}$ is an effective charge that couples to the flux of $\cal Z_\mu$. Taking this into account, we arrive to the Aharonov-Bohm phase (\ref{AB phase}).  In fact, Eqs.~(\ref{Gausss law A}) and (\ref{Gausss law Z})  give us a working definition for the charges of both the ${\cal A}_\mu$ and ${\cal Z}_\mu$ fields. The first equation gives $Q_{\cal A}=2\mu_1 \Phi_{\cal Z}$, while the second  gives $0=-Q_{\cal Z}+2\mu_1 \Phi_{B_{\cal A}}+2\mu_2\Phi_{B_{\cal Z}}$. Both of these relations can be checked against their definitions given in Sec.~\ref{sec:flux-charge}. 

Notice that according to this definition, ${\cal Z}_\mu$ does not carry a charge, which is expected since it is a short range field and its charge is screened. Now, in order for any number of vortices to have a zero net ${\cal A}_\mu$-charge, we must have $\sum_in_i=0$ for the winding numbers $n_i$; violating this condition will mean that the system has a logarithmically divergent energy. In other words, any collection of vortices will be confined if and only if it has a zero net charge. Therefore, our vortices are the dynamical realization of the Cornalba-Propitius-Wilczek classical confinement phenomenon \cite{Cornalba:1997gh,deWildPropitius:1997wu}.

In this work we did not discuss the stability of our vortices as it is beyond the scope of this paper. However, we expect the ones with $n= \pm 1$ to be stable against decay. As we discussed above, the Abrikosov-Nielsen-Olesen and global vortices lie on the opposite sides of the interval $\mu_1 \in [0,\infty]$ (Fig.~\ref{fig:parameter space}). The stability of both kinds of vortices were studied in \cite{Goodband:1995rt}, and it was found that the ones with $n=1$ are stable, as expected on topological grounds. Indeed, an analysis that follows the lines of \cite{Goodband:1995rt}  should be repeated for our vortices to insure their stability. However, since $n=1$ vortices on the boundaries of the $\mu_1$ interval were found to be stable, it is implausible that they  lose stability in between as we vary $\mu_1$.

\acknowledgements
We would like to thank G. Dunne, A.J. Long, K.D. Olum, E. Poppitz, T. Sulejmanpasic, T. Vachaspati and A.Vilenkin for useful conversations. This work is supported by the Swiss National Science Foundation. Y.B. is supported by the grant PZ00P2-142524.

\bibliographystyle{apsrev4-1}

\end{document}